# Long-lived and disorder-free charge transfer states enable endothermic charge separation in efficient non-fullerene organic solar cells

Philip C.Y. Chow[1,3]\*, Ture F. Hinrichsen[2]\*, Christopher C.S. Chan[1], David Paleček[2], Alexander Gillett[2], Shangshang Chen[1,3], Xinhui Zou[1,3], Chao Ma[1], Guichuan Zhang[4], Hin-Lap Yip[4], Kam Sing Wong[1], Richard H. Friend[2], He Yan[1,3], Akshay Rao[2]

Affiliations:

1. Department of Chemistry, Department of Physics and Energy Institute, The Hong Kong University of Science and Technology, Clear Water Bay, Hong Kong

2. Cavendish Laboratory, University of Cambridge, Cambridge, United Kingdom

3. HKUST Shenzhen Research Institute, Shenzhen, China

4. Institute of Polymer Optoelectronic Materials and Devices, State Key Laboratory of Luminescent Materials and Devices, South China University of Technology, Guangzhou 510640, China

\* These authors contributed equally to this work.

Correspondence: PCYC <philip.cy.chow@gmail.com;pcyc@ust.hk>, AR <ar525@cam.ac.uk>, HY <hyan@ust.hk>, or RHF <rhf10@cam.ac.uk>

**Abstract**

Organic solar cells (OSCs) based on non-fullerene acceptors can show high charge generation yields despite near-zero donor-acceptor energy offsets to drive charge separation and overcome the mutual Coulomb attraction between electron and hole. Here we use time-resolved optical spectroscopy to show that free charges in these systems are generated by thermally activated dissociation of interfacial charge-transfer excitons (CTEs) that occurs over hundreds of picoseconds at room temperature, three orders of magnitude slower than comparable fullerene-based systems. Upon free electron-hole encounters at later times, CTEs and emissive excitons are regenerated, thus setting up an equilibrium between excitons, CTEs and free charges. This endothermic charge separation process enables these systems to operate close to quasi-thermodynamic equilibrium conditions with no requirement for energy offsets to drive charge separation and achieve greatly suppressed non-radiative recombination.

**Main text**

Organic solar cells (OSCs) are flexible, semi-transparent and environmental-friendly alternatives to inorganic solar cells. In contrast to inorganic semiconductors, photoexcitation of organic semiconductors (molecules and polymers) generates coulombically-bound electron-hole pairs known as excitons [1]. These excitons have large binding energies (typically ~0.5eV) and do not separate into free charges unless they are dissociated at a donor/acceptor (D/A) heterojunction to form charge-transfer excitons (CTEs), as shown in Fig. 1a,b, which must subsequently separate to free electrons and holes against their mutual Coulomb interaction in order to create photocurrent. Bulk-heterojunction OSC blends with nanoscale morphology are able to achieve nearly 100% free charge generation yields [2]. However, their overall power conversion efficiency (PCE) is limited by the relatively low open-circuit voltage ($V_{OC}$) with respect to their optical gap ($E_{S1}$). Current models of operation in OSCs introduce three sources of photovoltage (energy) losses [3–6]. The first is losses due to photon entropy increase, which are unavoidable in all types of solar cells (typically ~0.3 V in the Shockley-Queisser framework) [7]. The second is non-radiative recombination of charge pairs, either by phonon mediated relaxation to the ground state [8] or formation of low-energy spin-triplet excitons which subsequently relax to the ground state [9,10]. The magnitude of the non-radiative voltage loss is given by $V_{nonrad} = -kT\,ln(EQE_{EL})/q$, where $k$ is the Boltzmann constant, $T$ the absolute temperature, $q$ the elementary charge, and $EQE_{EL}$ the electroluminescence quantum efficiency [11–13]. The third source of photovoltage loss is the offset between the energy of the photogenerated singlet excitons ($E_{S1}$) and the interfacial CTEs ($E_{CTE}$), see Fig. 1a. This energy offset, $E_{S1}$–$E_{CTE}$, is widely believed to be required in order to drive efficient and rapid (within a few hundred femtoseconds) charge separation at organic heterojunctions [2,14,15].

Previous studies have shown that a large $E_{S1}$–$E_{CTE}$ offset (~200 meV and above) is typically required to drive charge separation in conventional OSCs that use fullerene acceptors (such as [6,6]-phenyl-$C_{61}$-butyric acid methyl ester or PCBM). This offset requirement in turn leads to the formation of CTEs with energy below the optical gap, as evidenced by sub-gap emission and absorption tails of the blends [16]. Additionally, CTEs in fullerene systems often introduce energetic disorder that broadens the density of states (Urbach energy typically ≥40 meV for fullerene blends compared to ~25-30 meV for pure organic materials) that can increase charge recombination [17]. We note that there exist several examples of fullerene OSCs that can achieve decent performance with small offsets [18–20]; however, most fullerene systems engineered with small offsets show reduced charge generation quantum yields [6,21,22]. Furthermore, fullerene systems typically give $EQE_{EL}$ ≤$10^{-6}$, resulting in a typical non-radiative recombination loss of >0.35 V [13]. Overall, the combination of these terms results in photovoltage losses in the range of ~0.8-1.0 V with respect to $E_{S1}/q$ for fullerene OSCs with large energy offsets [23]. Alternatively, many recent examples of OSC systems based on non-fullerene acceptors show high charge generation yields with negligible $E_{S1}$–$E_{CTE}$ offsets (≤0.05 eV; see Fig. 1b), and high $EQE_{EL}$ (≥$10^{-4}$) [24–29]. The combination of low $E_{S1}$–$E_{CTE}$ offsets and high $EQE_{EL}$ results in greatly reduced photovoltage losses (≤0.6 V), and state-of-the-art materials with broad spectral overlap with sunlight can now achieve over 15-16% PCE in single-junction devices [30,31]. Considerable effort has been directed towards exploring new materials within this class and to understand their charge dynamics [24,32–34]. However, the detailed mechanism for charge separation in these efficient non-fullerene systems with small energy offsets is not fully understood.

Here, we study the charge separation process in efficient non-fullerene OSC systems with small photovoltage losses and high charge generation quantum yields. First, we use two-pulse (pump-probe; PP) transient absorption spectroscopy to probe the conversion of photogenerated excitons to CTEs and free charges. Then, we use three-pulse (pump-push-probe; PPP) transient absorption to selectively monitor the population of CTEs confined at the donor-acceptor interface. The combination of these optical methods allows us to obtain a detailed overview of the excited state dynamics as a function of time and temperature, revealing how free charges efficiently separate with suppressed non-radiative recombination losses in non-fullerene OSC systems with small energy offsets.

Our study involves four model non-fullerene blends, namely P3TEA:SF-PDI$_2$ [27], P3TEA:FTTB-PDI$_4$ [35], P3TAE:SF-PDI$_2$ [36] and PffBT2T-TT:O-IDTBR [37]. The chemical structures of these materials are shown in Fig. 1d, and their energy levels, absorption spectra, device performance and morphology characterisation can be found in Fig. S2-4. Among these we focus on the P3TEA:SF-PDI$_2$ blend. With an $E_{S1}$ of 1.72 eV (determined by the intercept of absorption and emission spectra), the large $V_{OC}$ of 1.11 V indicates a photovoltage loss of 0.61 V. Of this only ~0.25 V is due to non-radiative recombination losses as determined by the EQE$_{EL}$ of 0.5x10$^{-4}$. Fitting of the absorption tail and EL spectrum places the CTE energy at ~1.7 eV (see Fig. S4), indicating a negligible energy offset between $E_{S1}$ and $E_{CTE}$, as shown in Fig. 1b. The steep absorption edge shows that CTEs have low degree of electronic disorder (Urbach energy of ~27 meV). Despite the small $E_{S1}$-$E_{CTE}$ offset, P3TEA:SF-PDI$_2$ devices exhibit photovoltaic internal quantum efficiency (IQE) of around 85-90% at short-circuit, and therefore represent one of the most efficient OSC systems with small photovoltage loss to date. We note that the relatively modest PCE of 9.5% compared to other state-of-the-art systems is mainly due to its weak absorption of sunlight in the near-infrared region (see Fig. 1c). For comparison, we contrast this system with its less efficient fullerene-based counterpart: P3TEA:PCBM (PCE of 7.5%, with a photovoltage loss of 0.83V). The CTEs in the fullerene blend has energy of ~1.6 eV (see Fig. S4), giving an $E_{S1}$–$E_{CTE}$ offset of about 100 meV, and have low electronic disorder comparable to those found in the non-fullerene blend. We observe strong photoluminescence (PL) quenching for both blends with respect to the pure polymer film, thus indicating efficient dissociation of excitons into charges at D/A interfaces (see Fig. 4a). Negligible $E_{S1}$–$E_{CTE}$ offsets and low CTE disorder are also found in P3TEA:FTTB-PDI$_4$, P3TAE:SF-PDI$_2$ and PffBT2T-TT:O-IDTBR blends. Devices based on these blends can achieve large $V_{OC}$ of 1.13 V, 1.19 V and 1.08 V, respectively. The combination of small voltage losses (<0.6 V) and high external quantum efficiencies (>60%) allows these systems to achieve high PCEs despite their weak absorption in the near-infrared (see Fig. S2). As can be seen from Fig. 1d, the various non-fullerene acceptors cover a range of different structural motifs and allow us to generalise the results of our study to any non-fullerene system with near zero energy offsets to drive charge separation.

We first probe excited state dynamics using pump-probe (PP) transient absorption spectroscopy, where a broadband white-light pulse probes the films following excitation with a pump pulse (spectrum shown in Fig. 1c). Figure 2a-c compare the PP spectra of pristine P3TEA, P3TEA:SF-PDI$_2$ and P3TEA:PCBM films near the absorption edge (670-780 nm) at various pump-probe time delays (non-normalised spectra over the full detection range can be found in Fig. S8,9). Photoexcitation of the pristine polymer film leads to the formation of singlet excitons and we observe a positive ΔT/T signal up to ~740 nm that matches the polymer absorption (thus corresponding to the ground-state bleach; GSB). The PP response of the pristine polymer does not evolve spectrally up to 1 ns, indicating that singlet excitons are the only excited state species. At early times (0.2 ps; Fig. 2a), the non-fullerene blend (P3TEA:SF-PDI$_2$) shows a similar PP response to the pristine film, thus indicating that singlet excitons make up most of the excited states. In contrast, the fullerene (P3TEA:PCBM) blend already shows a negative ΔT/T signal emerging at ~720 nm onwards that is similar to the quasi-steady-state electroabsorption (EA) response measured in a diode structure. This negative ΔT/T signal is not found for the pristine polymer and is thus attributed to charges (polarons) created at the D/A interface from very early times [2]. More specifically, this negative ΔT/T signal can be explained by two partially overlapping features. We attribute the signal at longer wavelengths (>750 nm) to polaron absorption and the signal around 720 nm, which leads to a blue-shift of the absorption edge, to a transient EA response. As discussed in detail in previous work [2,14], the transient EA response arises from a Stark shift of the absorption spectrum due to the local electric fields generated between electron-hole pairs following charge separation at the D/A heterojunction. The emergence of this transient EA response in PP measurements is thus a signature of the charge separation process (see Supplementary Information for a detailed description). Our results indicate that the formation of free charges (spatially separated) occurs on an ultrafast timescale (<1 ps) in the fullerene blend, which is consistent with previous measurements on other fullerene-based systems with large D/A energy offsets.

With increasing time delay (Fig. 2b,c), a similar transient EA response emerges in the non-fullerene sample at ~20 ps, but does not evolve to match the response of the fullerene blend until ~200 ps. Remarkably, the timescale at which the transient EA response emerges for the non-fullerene system shows that long-range charge separation occurs at a much slower rate compared to the fullerene blend, taking over ~100 ps to complete. Similar charge separation times are also found in P3TEA:FTTB-PDI$_4$, P3TAE:SF-PDI$_2$ and PffBT2T-TT:O-IDTBR (see Fig. 3d-f). We highlight that by tracking the growth of the transient EA signal we monitor the separation dynamics of excitons/CTEs into free charges, while other reports to date focused on measuring the charge transfer time across the D/A heterojunctions (i.e. dissociation of excitons into CTEs) in non-fullerene OSCs [32].

The slow timescale for charge separation of CTEs measured here (~100 ps) contrasts with the fast rate of vibrational relaxation for excitons and CTEs (~100 fs) [23,38,39]. This means that CTE separation must occur from thermally-relaxed CTEs [40]. We note that no 'excess energy' from the donor-acceptor offset is available to drive charge separation for these systems (negligible $E_{S1}$–$E_{CTE}$ offsets). However, as the internal energy of free charges must lie above the coulombically-bound CTE state (see Fig. 1b), it follows that charges must overcome the Coulomb energy barrier to separate into free charges. This barrier, the CTE binding energy, is typically found to be 200-250 meV in organic blends [1,2,17,41]. We thus propose that the slow charge separation measured here for the non-fullerene systems is due to the need for thermal activation of the CTEs to free charges, making the process endothermic.

To explore this further we perform PP spectroscopy at reduced temperatures. For pristine P3TEA, we find a slight red shift (~<10 nm) of the GSB due to lowering of the optical gap upon cooling (see Fig. S10). For the non-fullerene blend at early times (~1 ps; inset of Fig. 2d), the PP response (associated mostly with singlet excitons) is largely unchanged with respect to that of the pristine polymer. However, as shown in Fig. 2d, we do observe a clear effect of temperature on the PP response at later times (~200 ps), showing much weaker signs of transient EA signal created by charge separation at reduced temperatures (as evidenced by the much reduced spectral blue-shift). This indicates that long-range charge separation in the non-fullerene blend is suppressed at lower temperatures, providing evidence that it is indeed an endothermic process. In contrast, reducing temperature has insignificant effect on long-range charge separation in the fullerene blend, where a clear EA signal is seen even at ~20 K (see Fig. S10). This result is consistent with the considerable $E_{S1}$–$E_{CTE}$ offset measured for this system (see Fig. 1a and Fig. S4) and agrees with previous reports that charge separation via the delocalised electronic states of fullerene acceptors is a temperature-independent process [2,42].

When examining the effect of temperature on the charge generation yield, we find that almost no photocurrent is generated below ~120 K for the P3TEA:SF-PDI$_2$ device (see Fig. S5). In comparison, we find that P3TEA:PCBM has a much weaker dependence on temperature. Weak temperature dependence of photocurrent yield is also found in other efficient fullerene-based systems that are designed with energy offsets to drive charge separation [42]. These results are consistent with the proposed endothermic nature of free charge generation in the non-fullerene blends studied here. We note that the activation energy determined by fitting of the steady-state photocurrent at various temperatures does not provide a direct measure of the Coulomb binding energy of CTEs, but instead provides a measure of the interplay between charge dissociation and recombination rates at various temperatures.

While PP spectroscopy allows us to track the spatial separation of charges via growth of the transient EA signal, it does not provide a picture of the size of CTE population confined at the heterojunctions. To observe this, we turn to pump-push-probe (PPP) spectroscopy. In this technique the sample is excited twice: an above gap 'pump' pulse generates an excited state population, which is then influenced by a 'push' pulse. The push wavelength is selected so that it is not absorbed by the ground state, as shown in Fig. 1c, but can be absorbed by excited states (excitons, CTEs and free charges). The optical response to this perturbation is then recorded by a third, broadband probe pulse. For the P3TEA:SF-PDI$_2$ blend, absorption of the 'push' by excitons leads to annihilation of excited states, whereas absorption by free holes causes no change in the spectrum. However, when absorbed by CTE states, the push causes an increase in electron-hole separation, as illustrated in Fig. 3a, which we observe as an increase in the EA response [17,43]. Figure 3b shows the PPP response of P3TEA:SF-PDI$_2$ for several pump-push delay times with a constant push-probe delay of 1 ps. Here, we have subtracted the 'annihilation' component from exciton absorption to reveal the effect of the push pulse on the CTE states (see Supplementary Information for details on this procedure). Also shown is the quasi-steady-state EA response measured in a diode structure which matches the PPP spectrum near the absorption edge (~700 nm), showing that the push pulse indeed separates CTE states and gives rise to an EA response. By tracking the strength of this PPP response, we are thus able to monitor the population of coulombically-bound CTEs confined at the heterojunctions.

Figure 3c shows the evolution of this PPP spectral response in the P3TEA:SF-PDI$_2$ blend. We observe a drop in CTE population between ~10-100 ps, which agrees with the PP data (Fig. 2) that show long-range charge separation occurring on this timescale at room temperature. However, the CTE population does not fall to zero, with almost half of the initial CTE signal still present at 800 ps. From the PP data, we know that there is no growth in EA beyond 200 ps, indicating that the population of free charges does not substantially increase beyond this point. We know that free charge generation is very efficient in this blend, with IQE of ~85-90%, so that the long-time CTE signal is associated with CTEs that do eventually separate to free charges. We have noted above that the slow charge separation we observe here occurs from thermalized CTEs, which will also be the states populated via bimolecular encounters of free electrons and holes. We consider therefore that this long-lived PPP signal arises from CTE formed at long-timescales due to bimolecular charge encounters. Thus, we observe here the build-up of a quasi-equilibrium between free charges and CTEs, in agreement with previous suggestions in the literature [33,41].

At lower temperatures we find the PPP response in the P3TEA:SF-PDI$_2$ blend increases, indicating an increase in CTE population. The inset of Fig. 3c shows the CTE signal at 50-100 ps after the initial photoexcitation at various temperatures, normalised to the singlet exciton signal. The corresponding spectra are shown in Fig. S12. This indicates that the temperature dependence of charge separation observed in PP measurements (Fig. 2) is not due to trapping of excitons caused by morphology change at reduced temperature, but instead due to the endothermic nature of CTE separation in the non-fullerene system. Tracking the relative CTE population, Fig. 3c, after the initial charge separation has occurred, we find that the population of bound CTE states decreases by only 30% by 200 ps. This suggests that within these systems the quasi-equilibrium favours the bound CTE states in the absence of charge extraction, i.e. open-circuit conditions. Without a thermally activated charge separation channel (which we observe both after initial photoexcitation and following free electron-hole encounters), this would lead to poor device IQEs.

For P3TAE:SF-PDI$_2$, P3TEA:FTTB-PDI$_4$ and PffBT2T-TT:O-IDTBR, we also observe a long-lived PPP spectral response with derivative line-shape near the optical edge that we attribute to CTEs at the D/A heterojunction of these blends (see Fig. 3d-f). These CTE signals do not fall to zero beyond the free charge generation timescale (up to 800 ps; see Fig. S14), in agreement with our observations for the P3TEA:SF-PDI$_2$ blend of a quasi-equilibrium between CTEs and free charges. Overall, we observe very similar PP and PPP results in the four efficient non-fullerene blends with small photovoltage losses despite their different structural motifs, thus showing that the slow (tens to hundreds of ps) yet efficient endothermic charge separation process is general to non-fullerene systems with near-zero $E_{S1}$–$E_{CTE}$ offsets. In contrast, we observe a very weak PPP response in P3TEA:PCBM at ~200 ps (see Fig. S13), thus indicating that most charges are spatially separated and few CTEs are present.

We now turn to the charge recombination dynamics. As shown in Fig. 4a,c, the PL spectral shapes of pristine P3TEA and P3TEA:SF-PDI$_2$ films are similar despite the large difference in intensity (PL quantum yield of ~2% and ~0.2%, respectively). Figure 4b shows time-resolved PL measurements, where following the initial PL quenching due to dissociation of excitons, we observe a delayed emission on a nanosecond timescale for the non-fullerene systems. Importantly, this delayed emission is longer lived than the PL of the pristine material and is not observed for the fullerene blend (PL quantum yield of ~0.1%). This suggests that it is due to radiative recombination of long-lived CTEs found in the non-fullerene blend.

We further study the PL of these materials in a diode structure. As shown in Fig. 4c, we find that the PL intensity of the non-fullerene blend decreases uniformly with applied reverse bias, whereas no difference is observed for the pristine polymer device over the same bias range (see inset). This indicates that, while the applied electric field is insufficient for separating excitons in the pure material, in the blend it causes the free electrons and holes to drift towards opposite electrodes and thus reduces the radiative charge recombination [44]. Therefore, we obtain the emission spectrum arising from charge recombination by taking the differential PL spectrum of the device (PL at short-circuit minus PL with reverse bias of 10 V). We find that this differential emission closely resembles the PL spectrum of the pristine material. Since the CTE energy is close to that of the singlet exciton of the donor polymer (Fig. 1b), it is very likely that the emission caused by charge recombination is from singlet excitons that are populated by electron back-transfer from CTEs [32,45,46]. Furthermore, as shown in Fig. 4d, we find that the application of reverse bias suppresses the aforementioned delayed emission observed in the non-fullerene blend (consistent with the reduced PL intensity).

We thus assign this bias-dependent, nanosecond PL to singlet excitons formed via CTEs that are regenerated upon bimolecular encounters of free electrons and holes at the D/A heterojunction. This is consistent with the response observed at longer times in the PPP measurements (Fig. 3) and implies that the quasi-equilibrium includes the singlets excitons as well as CTEs and free charges, allowing for excitations to cycle between the states until they recombine to the ground state via the singlet exciton state (CTE to ground state transition is much slower [32]). This indicates that the charge separation yield can be limited by fast singlet recombination [46], and thus explains the correlation between open-circuit voltage and PL quantum yield of the low-gap component found in other non-fullerene systems with small offsets [32]. Since CTEs take ~100 ps to fully separate at room temperature, the singlet exciton lifetime must exceed this timescale for efficient charge separation to occur (this is consistent with the effective PL quenching found for P3TEA:SF-PDI$_2$ despite the regeneration of singlet excitons). The presence of an equilibrium between excitons and charges also implies that the singlet regeneration rate is comparable to the rate at which singlets form CTEs (~1-20 ps according to our PP results, see Fig. 2). We note that we find no evidence for bimolecular recombination leading to triplet exciton formation in P3TEA:SF-PDI$_2$, in contrast to the fullerene blend, thereby eliminating another non-radiative loss pathway (see Fig. S11 for further discussion).

For conventional fullerene-based OSCs with large offsets it has been widely noted that the rate of bimolecular recombination is suppressed compared to Langevin recombination models, and a number of explanations such as phase segregation of electrons and holes have been put forward to explain this [41,47]. However, in these fullerene systems, free electron-hole encounters do not regenerate singlet excitons to any significant extent as there is a substantial energetic offset between the relaxed CTEs (which are formed via bimolecular charge encounters) and the singlet excitons (Fig. 1a). The large energetic offset in turn is required to drive rapid charge separation. This implies that separation of the relaxed CTEs is not efficient in most systems involving fullerene acceptors, as reduced offset generally leads to lower EQE [6,21,22]. A likely reason for this is the high degree of electronic disorder found in many fullerene blends (Urbach energy $\geq 40$ meV) which traps the CTEs in low energy sites, from which they cannot escape to form free charges or regenerate singlet excitons due to fast non-radiative recombination [8,48]. Fullerene systems with small offsets that do show decent performance were found to have low electronic disorder (for example, PIPCP:PCBM [17] and PNOz4T:PCBM [32]). Thus, due to the requirement of an energetic offset for efficient charge separation in most fullerene-based systems, the $EQE_{EL}$ values for fullerene OSCs are generally very low ($\leq 10^{-6}$, as measured here for P3TEA:PCBM system).

However, our results presented here indicate that in the non-fullerene systems with small energetic offsets charge separation occurs at up to hundreds of picoseconds from thermalized CTEs via thermal activation. These same CTEs are repopulated via free electron-hole encounters, allowing for the regeneration of singlet excitons, which consequently establishes an equilibrium between the singlet excitons, CTEs and free charges. We note that such efficient separation of the relaxed CTEs in the non-fullerene systems is likely assisted by the low degree of electronic disorder at their interfaces, which is comparable to thermal energy at room temperature (Urbach energy ~27 meV in P3TEA:SF-PDI$_2$, see Fig. S4). This much more efficient regeneration of emissive singlet excitons in the lower gap material (in this case, P3TEA) enables a higher EQE$_{EL}$ ( ≥10$^{-4}$ for the non-fullerene systems with small offsets [29,32,49]) and hence lower non-radiative voltage losses. Importantly, the regenerated singlet excitons can then once again form CTEs that have a high probability of re-dissociating into free charges at room temperature.

The reversible interconversion of free charges, CTEs and singlet excitons suggest that their Gibbs free energies must be very similar for the non-fullerene OSC systems. This contrasts with conventional fullerene-based systems, where the large D/A energy offset significantly lowers the Gibbs free energy of the free charges with respect to CTEs and singlet excitons, thus introducing irreversibility into the system. This makes conventional fullerene OSC systems similar to natural light harvesting complexes, in which an energy cascade is also used to drive long-range charge separation on ultrafast timescales. This exothermic process leads to near-unity IQE but only in the expense of a large photovoltage loss. However, as we have shown here, highly efficient endothermic charge separation can occur in the non-fullerene blends via thermal activation of long-lived and disorder-free CTEs on long timescales (>100 ps). This endothermic process removes the need for charges to suffer a loss in free energy in order to obtain near-unity IQE and allows for minimal photovoltage loss.

Efficient endothermic charge separation brings non-fullerene OSCs, in terms of their thermodynamics, into the same mode of operation as inorganic solar cells, where encounters of free carriers do not directly lead to terminal recombination (in the absence of non-radiative decay events associated with defect states) and the Gibbs free energy of the photogenerated electron-hole pair and separated electron and hole are very close. These results reveal that OSCs are thus not limited by the Coulomb energies that bind excitons and CTEs, due to the possibility of undergoing thermally activated charge separation. Future OSCs should be designed to remove all irreversible processes and energy offsets, with significant efficiency gains being possible by reducing non-radiative losses in recombination. This could be achieved for example via the use of donor and acceptor materials with high PL quantum yields such that regeneration of excitons via free electron-hole encounters primarily leads to radiative recombination. It is also now time to address issues such as quenching of PL at charge collection electrodes, much investigated for lead halide perovskite PV systems, that have hitherto not been regarded as critical to performance for OSCs. This will enable OSCs to achieve high charge generation efficiencies with small photovoltage losses that are comparable to non-excitonic solar cells based on inorganic and hybrid semiconductors [30,31].

**Supplementary Information**

Materials and Methods

Supplementary Text

Fig. S1 to S14

**Acknowledgements**

P.C.Y.C., S.C., X.Z. and H.Y. acknowledge the Hong Kong Research Grant Council (project no. 16306117, 16304218, 16306319, R6021-18, C6023-19, 16305915, 16322416, 606012, 16303917) and the Shenzhen Technology and Innovation Commission (project no. JCYJ20170413173814007 and JCYJ20170818113905024) for funding support. K.S.W., C.C.S.C and C.M. acknowledges Hong Kong Research Grant Council (AoE/P-02/12). H.L.Y. acknowledges the Ministry of Science and Technology for funding support (no. 2017YF0206600 and no. 2019YFA0705900). The authors thank Matt Menke and Alexandre Cheminal for discussions.

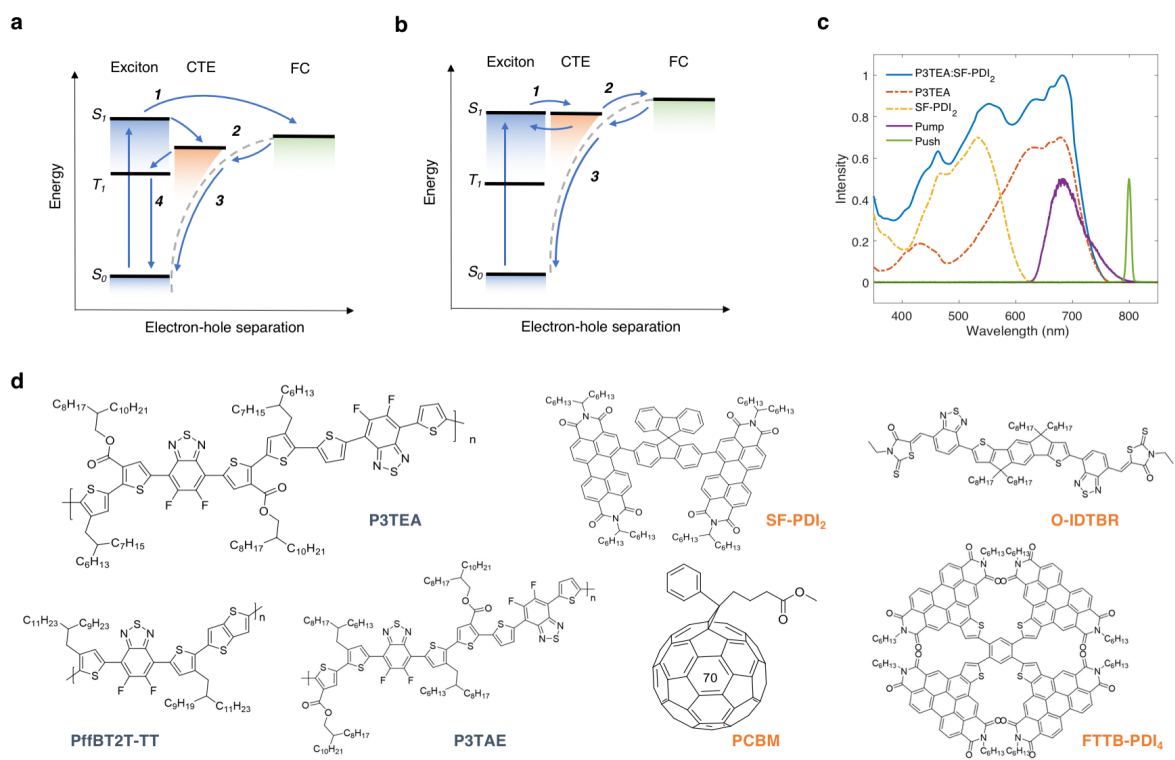

**Figure 1. Understanding charge separation processes in highly efficient non-fullerene organic solar cell (OSC) with small photovoltage loss.** (**a,b**) Schematic energy diagram illustrating the time evolution of charges in (**a**) conventional fullerene and (**b**) non-fullerene OSC systems. Conventional fullerene systems are designed with large donor-acceptor (D/A) offsets and generally have high degree of electronic disorder at the interface (Urbach energy ≥40 meV), whereas non-fullerene systems are designed with small offsets and generally have low disorder (Urbach energy ~25-30 meV). Photoexcited states evolve through three stages, 1: dissociation of singlet excitons into CTEs or directly into free charges, 2: separation of CTEs into free charges, 3: charge recombination. Additional losses through triplet exciton formation is often found for fullerene systems (process 4). (**c**) Absorption spectra of pristine P3TEA and SF-PDI$_2$, and P3TEA:SF-PDI$_2$ blend on quartz substrate. We preferentially create photoexcitations in P3TEA (pump at 670 ± 40nm) for all sub-nanosecond transient absorption measurements presented in this work. For pump-push-probe measurements, a push pulse (time-

delayed to the pump) arrives to interact with excited states and create a characteristic optical response that is subsequently monitored by a broadband probe pulse. The push pulse has an energy below the optical gap (push energy = 800 ± 10nm) and does not generate more photoexcitations directly from the ground state (see Supplementary Information). (**d**) Chemical structure of donor polymers (P3TEA, P3TAE and PffBT2T-TT) and acceptor molecules (SF-PDI$_2$, O-IDTBR, FTTB-PDI$_4$, and PCBM) involved in this study. Additional material and device characterisations are found in the Supplementary Information.

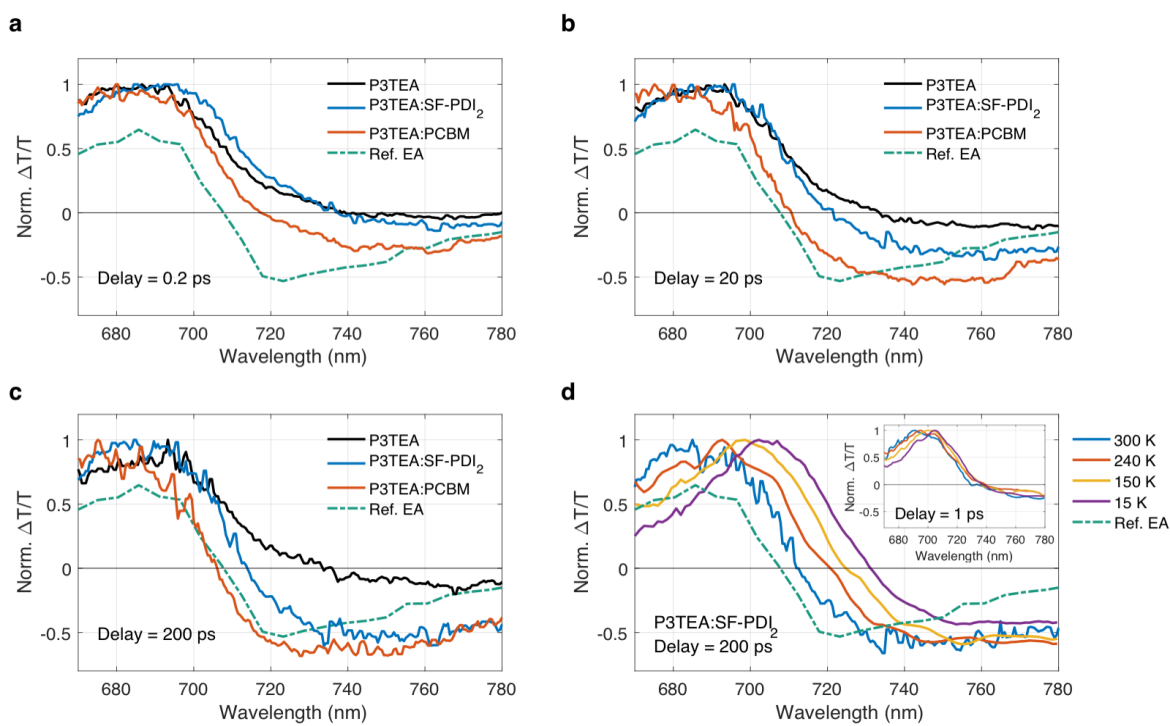

**Figure 2. Monitoring charge separation dynamics using pump-probe (PP) spectroscopy.** (**a,b,c**) Normalised PP transient absorption response (ΔT/T) for pristine P3TEA, P3TEA:SF-PDI$_2$ and P3TEA:PCBM near the absorption edge at room temperature (photoexcited at 670 nm, 1.5 μJ cm$^{-2}$ per pulse). Non-normalised data over full detection range can be found in Fig. S8,9. At early times (0.2 ps; panel **a**), the non-fullerene blend shows similar PP response to the pristine polymer which is associated with photogenerated excitons, while a negative ΔT/T signal from ~720 nm onwards associated with charges (hole polarons) has already emerged for the fullerene blend. As detailed in the main text, this negative signal is due to the growth of a transient electroabsorption (EA) signal caused by charge separation across the D/A interface as well as polaron absorption. With increasing time delay (20 ps; panel **b**), a similar negative ΔT/T signal emerges in the non-fullerene blend, but not evolving to match the fullerene ΔT/T signal until ~200 ps (panel **c**) at which both signals are similar to the quasi-steady-state EA

response measured in a diode structure (Ref. EA). This indicates that charge separation in the non-fullerene blend takes much longer to complete compared to the fullerene blend. (**d**) PP response of P3TEA:SF-PDI$_2$ at reduced temperatures. At early time (1ps; inset) the response (associated mostly with singlet excitons) is largely unchanged with respect to that of the pristine polymer (see Fig. S10). We note that the slight spectral red-shift (~<10 nm) of the ground-state bleach with reduced temperature is due to lowering of the optical gap. At later times (~200 ps), we find much weaker signs of transient EA signal created by charge separation at reduced temperatures (as evidenced by the much reduced spectral blue-shift), thus implying that it is indeed an endothermic process. In contrast, reducing temperature has insignificant effect on charge separation in the fullerene blend, where a clear EA signal is seen even at ~20 K (see Fig. S10).

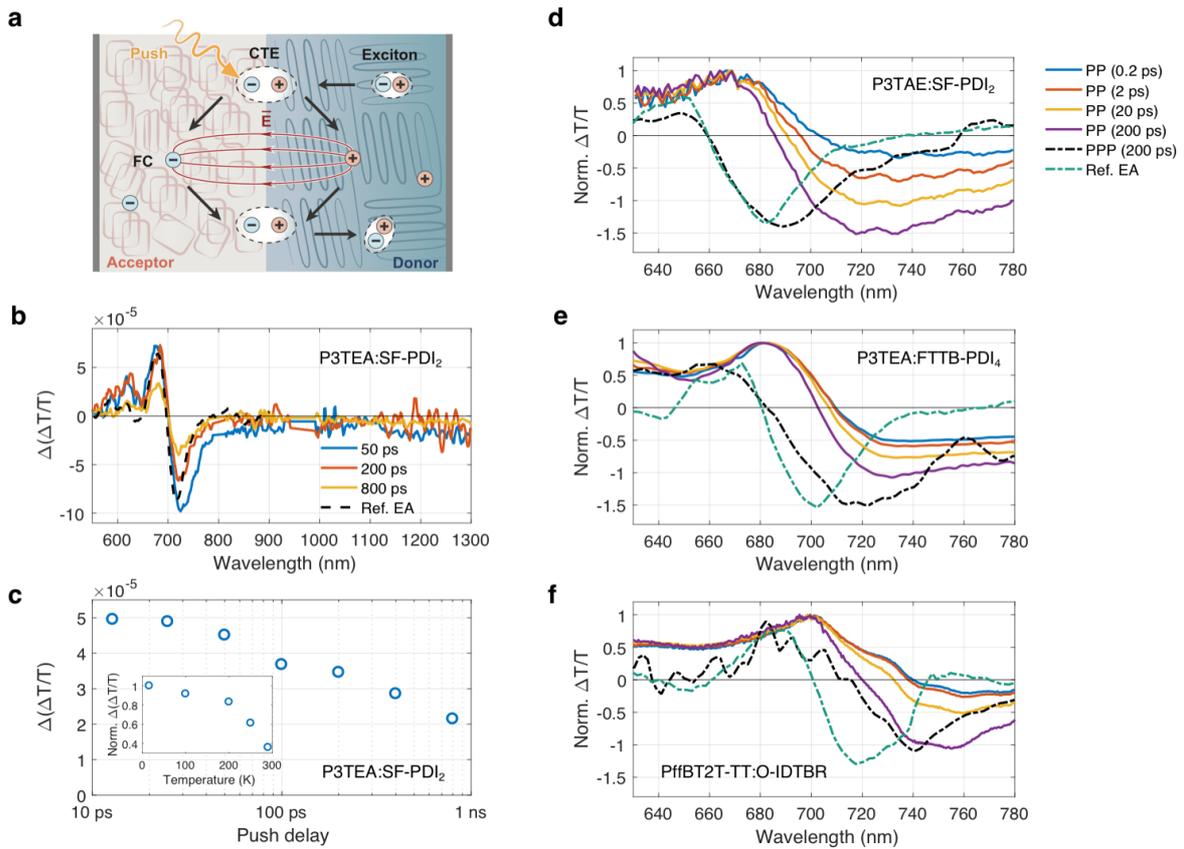

**Figure 3. Tracking charge transfer excitons (CTE) using pump-push-probe (PPP) spectroscopy.** (**a**) Schematic diagram of the excited state dynamics at the donor-acceptor heterojunction. If the push is absorbed by a CTE state it increases charge separation, causing an electroabsorption (EA) response near the absorption edge. (**b**) Change in transient absorption response (Δ(ΔT/T)) for P3TEA:SF-PDI$_2$ measured at various pump-push time delays (after subtraction of effects due to annihilation of excitons, see Supplementary Information). The signal near the absorption edge (~700 nm) closely resembles the quasi-steady-state EA response measured in a diode structure (Ref. EA) and can therefore be attributed to the push-induced response of bound CTE population. The pump (670 nm) and push (800 nm) pulses have a fluence of 50 μJ cm$^{-2}$ and 35 μJ cm$^{-2}$ per pulse, respectively. (**c**) Evolution of the CTE population (absolute value of signal between 640 and 740 nm), showing

remarkably slow charge generation. The inset shows the temperature dependence of CTE signal per photogenerated singlet exciton at 50-100 ps. At lower temperatures the CTE population increases, consistent with endothermic charge separation. (**d,e,f**) Pump-probe (PP) and pump-push-probe (PPP) spectra of P3TAE:SF-PDI$_2$, P3TEA:FTTB-PDI$_4$, and PffBT2T-TT:O-IDTBR (sharing the same legend). All blends show a growing charge signal in the PP spectra (negative ΔT/T response near the absorption edge) and an EA response in the PPP spectra at 200 ps (compared here to the derivative of the PP signal of the pristine polymer at 1 ps; labelled as Ref. EA).

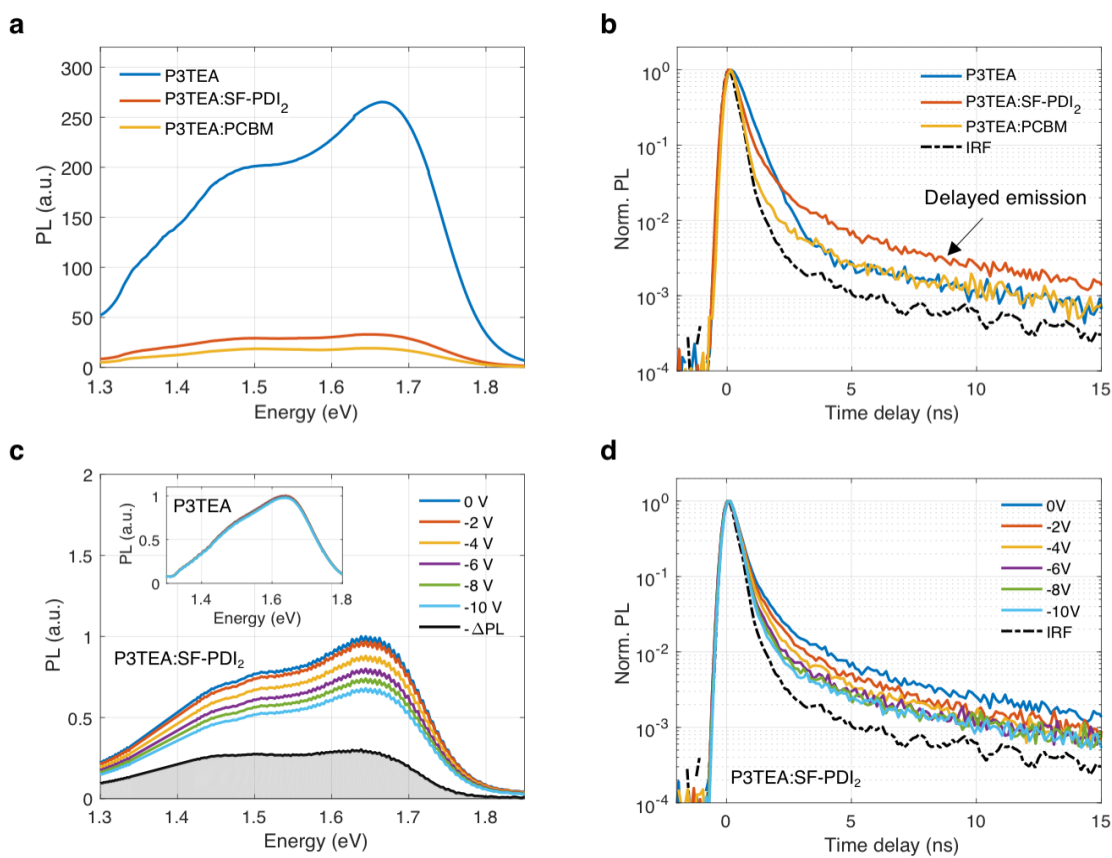

**Figure 4. Regeneration of emissive singlet exciton states upon free electron-hole encounters.** (**a**) Time-integrated PL spectra of pure P3TEA, P3TEA:SF-PDI$_2$ and P3TEA:PCBM thin-films under 633 nm excitation (corrected for difference in absorbance between samples). Using an integrating sphere we estimated a PL quantum yield of ~2% for the pure polymer film, ~0.2% for the non-fullerene film and ~0.1% for the fullerene film. (**b**) Time-resolved PL decay of thin-film samples under pulsed 680 nm excitations. A 780 nm (1.58 eV) long-pass filter was used to define the detection wavelength range. Additional measurements with a specific detection wavelength at ~750 nm (1.65 eV) show similar decay kinetics for the non-fullerene sample (see Fig. S6). Initial PL quenching at early times reflect efficient exciton dissociation in both blends, but delayed emission on a nanosecond timescale is observed only for the non-fullerene blend (consistent with the higher PL quantum yield). (**c**) Time-integrated PL of P3TEA:SF-PDI$_2$ device at various applied bias conditions, showing reduced PL with increasing reverse bias. The black line -ΔPL (PL at 0 V minus PL at -10 V) corresponds to the emission

that arises from charge recombination events (i.e. reverse bias suppresses free electron-hole encounters). The inset shows PL of a pure P3TEA device taken over the same bias range, showing negligible differences. (**d**) Time-resolved PL decay of P3TEA:SF-PDI$_2$ under various bias conditions (same excitation and detection conditions as in panel **b**). The reduction of delayed emission by applied reverse bias indicates that emissive singlet states are regenerated upon free electron-hole encounters. IRF: instrument response function.

# Long-lived and disorder-free charge transfer states enable endothermic charge separation in efficient non-fullerene organic solar cells

**Supplementary Information**

Materials and Methods
Supplementary Text
Figs. S1 to S14

**Materials and Methods**

Thin-film sample preparation

P3TEA, P3TAE, PffBT2T-TT, SF-PDI$_2$, FTTB-PDI$_4$ and O-IDTBR were synthesized as reported in Ref. [27,35-37] of main text, and PCBM was purchased from 1-Materials. The P3TEA:SF-PDI$_2$ and P3TEA:PCBM (both 1:1.5 w/w, polymer concentration 9 mg ml$^{-1}$) blends and pristine P3TEA films were prepared in 1,2,4-trimethylbenzene (TMB) with 2.5% of 1,8-octanedithiol (ODT) and left for stirring at 100 °C for at least 1 hour prior to film deposition. Quartz substrates were cleaned by sonicating in deionized water, acetone, and isopropanol for 5 minutes successively, followed by oxygen plasma treatment for 10 min. Thin-films were spin-coated from hot solutions on the preheated substrate in a N$_2$ filled glove box at 1500 r.p.m. to obtain thicknesses of ~120 nm and then thermally annealed for 10 minutes. All samples were encapsulated using epoxy and glass inside the glovebox to avoid degradation during measurements. Ultraviolet–visible absorption spectra were acquired using a Perkin Elmer Lambda 20 UV/VIS Spectrophotometer. GIWAXS measurements were carried out on a Xenocs Xeuss 2.0 system with an Excillum MetalJet-D2 X-ray source operated at 70.0 kV, 2.8570 mA, and a wavelength of 1.341 Å. The grazing-incidence angle was set at 0.20°. Scattering pattern was collected with a DECTRIS PILATUS3 R 1M area detector.

Device preparation and characterization

Solar cell devices were prepared by spin-coating the photoactive materials onto pre-cleaned ITO-coated glass with sheet resistance of 15Ω per square instead of quartz substrates. Then, a thin layer (20 nm) of V$_2$O$_5$ was deposited as the anode interlayer, followed by deposition of 100 nm of Al as the top electrode (at a vacuum level of $3 \times 10^{-6}$ torr). Each device pixel had an area of 5.9 mm$^2$ and was encapsulated with epoxy and glass. Device J–V characteristics were measured under AM1.5G (100 mWcm$^{-2}$) at room temperature using a Newport solar simulator, calibrated using a standard Si diode (with KG5 filter, purchased from PV Measurement) to bring spectral mismatch to unity, and a metal mask with an aperture aligned with the device area was used during measurements. EQE was characterized using a Newport EQE system.

Quasi-steady-state electroabsorption

The Quasi-steady-state electroabsorption (EA) spectrum was acquired by measuring the relative change in reflection off the back electrode *(ΔR/R)* at various wavelengths in response to a varying electric field applied across the electrodes of the device, with the beam passing through the donor-acceptor blend twice. The device was held under a small (-1V) reverse bias to minimize charge injection from the electrodes. Since this small bias does not affect the reflectance of the back electrode, the relative change in reflection is equivalent to the relative change of transmission *(ΔT/T)* through the sample.

Photoluminescence and electroluminescence spectroscopy

The time-integrated photoluminescence (PL) spectrum of thin-film and device samples was measured using a calibrated Andor iDus CCD camera. An integrating sphere was used to determine PL quantum yields. The device samples were connected to a Keithley 2400 source meter for studying the change in PL at reverse bias and for driving electroluminescence (EL) at forward bias.

For time-resolved PL, the encapsulated thin-film or device samples were photoexcited with a Fianium laser running at 5 MHz. A 680 nm band pass filter was used to select the excitation wavelength. The collected PL was measured with an InGaAs time-correlated single photon counter (TCSPC). Long-pass and band-pass filters were used for selecting specific detection ranges.

Transient absorption spectroscopy

For sub-ns pump-probe measurements, both pump and probe pulses were generated in a home-built non-collinear optical parametric amplifier (NOPA) with the second harmonic of a Ti:Sapphire laser (Spectra Physics, 1kHz). An electronically delayed Q-switched Nd:YVO4 laser (532 nm, Advanced Optical Technologies Ltd.) was used to provide pump pulses for measurements beyond 1ns. For pump-push-probe measurements, all optical pulses were generated from a PHAROS (Light Conversion) laser running at 38 kHz. The probe pulse was a chirped white light continuum generated by focusing the fundamental (1030nm) in a YAG crystal. The 670 nm pump pulse was generated in a home-built non-collinear optical parametric amplifier, and the 800 nm push pulse was generated by feeding the PHAROS output into a commercial ORPHEUS (Light Conversion) optical parametric amplifier. The optical pulses were spatially overlapped in the sample through a boxcar geometry, and temporally delayed using a piezoelectric delay stage for the pump-probe delay and a DC servo stage for the pump-push delay. After passing the sample, the probe was split with a low pass filter with 960 nm cut-off to obtain a transmitted near infrared beam and a reflected visible beam. The near infrared and visible probe beams were guided into a spectrometer with an InGaAs and Si photodiode array detector, respectively. The two sensors were synchronized and detected simultaneously at 38 kHz, enabling shot-to-shot detection. Pump and push beams were chopped at 38/4 kHz and 38/8 kHz, respectively, such that we simultaneously recorded the pump-induced change in transmission ($\Delta T/T$) and the pump-push-induced response ($\Delta(\Delta T/T)$). The push-induced change in transmission in the absence of a pump (push-probe response) was also recorded, which allowed us to minimize multi-photon excitations of the ground state by the push. The pump and push fluences were set to obtain a good signal to noise for the pump-push-probe signal $\Delta(\Delta T/T)$ with minimum pump and push fluences, and the pump-probe signal measured was used to check for sample degradation.

**Supplementary Text**

**Section S1. Additional material and device characterizations**

1.1. UV-Vis absorption and photovoltaic performance

Fig. S2a,b show the absorption profile and energy levels of the materials involved in this study. The absorption data was measured in thin-films and the energy levels were estimated from cyclic voltammetry. Fig. S2c,d show the solar cell device performance of the five studied blends. The peak internal quantum efficiency (IQE) is above 85% for the model P3TEA:SF-PDI$_2$ device. Also summarised is the total voltage loss of each device, which is given by the difference between optical bandgap ($E_{gap}/q$) and the open-circuit voltage ($V_{OC}$). All non-fullerene systems studied herein exhibit much lower voltage losses compared to fullerene devices (typically ~0.8-1.0 V). We note that the value of voltage loss depends on how the optical gap is determined (see Wang et al. Adv. Energy Mater. 8, 1801352 (2018)). Here, all optical gaps were determined from the crossing point between absorption and emission spectra, as opposed to the onset of either absorption or emission spectra. We note that even smaller voltage losses (by about ~0.05 V) are calculated for each system using optical gaps taken from the absorption onset, and therefore the values quoted in the table below represent the upper bound of the voltage losses.

1.2. Morphology characterization

Detailed analysis of the blend morphology of P3TEA:SF-PDI$_2$, P3TAE:SF-PDI$_2$, PffBT2T-TT-O-IDTBR and P3TEA:FTTB-PDI$_4$ blends are described in Ref. [27,35-37] of main text. Using resonant soft-X-ray scattering (R-SOXS) and grazing-incidence wide-angle X-ray scattering (GIWAXS), it was found that all of these blends have similarly small domain sizes (~20-30nm) with high domain purity. Furthermore, by tracking the high-order lamellar stacking peaks, it was found that the donor polymers in these blends exhibit high crystallinity with preferential packing in the out-of-plane direction. Additionally, atomic force microscopy (AFM) measurements taken on these blended films show a uniform and smooth surface without clear phase separation (Fig. S3a). The morphologies of these blends are therefore very similar, showing a combination of fine intermixing between donor and acceptor materials and highly pure and crystalline domains which is well-known to be desirable for bulk-heterojunction organic solar cells. The similarity of morphologies in these blends is consistent with their similar charge generation dynamics that we reveal in this work. For P3TEA:SF-PDI$_2$ and P3TEA:PCBM, our GIWAXS data also show similar polymer packing in the out-of-plane direction (1.75 Å$^{-1}$), thus indicating that they have similar morphology as well (Fig. S3b). Further evidence for this is the significant PL quenching shown in both blends (Fig. 4), which implies that both blends have nanoscale domain allowing excitons to diffuse to a heterojunction and separate. These two blends also have similar energetic disorder at the D/A interface (see Section 1.3 for details).

1.3. Sub-gap absorption and emission spectroscopy

We used photo-thermal deflection spectroscopy (PDS) and electroluminescence (EL) spectroscopy to study sub-gap absorption and emission characteristics of P3TEA, P3TEA:SF-PDI$_2$ and P3TEA:PCBM. Fig. S4a shows the PDS data measured in thin films. We find that the three films show similar sub-gap absorption profiles, indicating that charge-transfer states formed at the donor-acceptor heterojunctions for the two blends have similar energies as the lowest singlet excitons. The steep absorption edges show Urbach energies of ~27 meV, as indicated by the dashed line. This value is considerably smaller than those previously measured for other organic heterojunctions, particularly those involving fullerene acceptors (typically ~40-50 meV). This provides evidence that the energy landscape at the donor-acceptor interface of these blends is relatively ordered and free of deep trap sites which would lead to fast non-radiative recombination Ref. [17] of main text.

Fig. S4b,c show the electroluminescence (EL) spectrum of the non-fullerene and fullerene blend, respectively. The coloured lines represent different injection biases (between 1.5 to 3V). We find almost no difference in the emission spectral shape with increasing injection bias for the non-fullerene blend (in agreement with previous report, see Ref. [27] of main text), while for the fullerene blend we find decreasing emission below ~1.5 eV with respect to the peak emission at ~1.65 eV. As described previously (Ref. [18] of main text), this spectral effect is observed in cases where the charge-transfer exciton states (CTEs) have lower energy compared to the singlet excitons. More charges recombine via the CTEs at small injection bias, while more excitons are populated at large injection bias. Therefore, our results indicate that for the non-fullerene blend the CTEs have close energy alignment with the singlet excitons, while for the fullerene blend the CTEs have lower energy than the singlets.

We quantify the CTE energy ($E_{CT}$) by simultaneously fitting the absorption edge and EL spectra using equations derived from Marcus theory (more details in *(18)*), and from that extract the band gap of the emitting state:

$$Abs_{pv}(E) = \frac{f}{E\sqrt{4\pi\lambda kT}} exp\left(\frac{-(E_{CT} + \lambda - E)^2}{4\lambda kT}\right)$$

$$EL_{pv}(E) = \frac{f}{E\sqrt{4\pi\lambda kT}} exp\left(\frac{-(E_{CT} - \lambda - E)^2}{4\lambda kT}\right)$$

where E is the photon energy, k is the Boltzmann constant, and T is the temperature. The fit parameters are $E_{CT}$, the reorganization energy $\lambda$, and a factor f proportional to the number of states. The fits are shown as the dashed lines. For P3TEA:SF-PDI$_2$, we fit the absorption and EL spectra at the peak since we observed no injection bias dependence on the EL spectrum, thus indicating CTEs have very similar energy as the lowest singlet excitons at 1.72 eV. For P3TEA:PCBM, our bias-dependent EL data indicates that CTEs have slightly lower energy than singlets, and their emission is red-shifted towards ~1.5 eV. As described by Ran et al. 10.1002/adma.201504417 (Ref. [18] of main text), it is possible to fit the low energy shoulder of the EL as the CTE shoulder. We therefore fit the low energy shoulder of the EL and we estimate an $E_{CT}$ of 1.62 eV for the fullerene blend. Similar data for P3TAE:SF-PDI$_2$, P3TEA:FTTB-PDI$_4$ and PffBT2T-TT:O-IDTBR are found in Ref. [35-37] of main text.

1.4. Temperature dependent photocurrent measurements

We measured the temperature-dependent photocurrent of P3TEA:SF-PDI$_2$ and P3TEA:PCBM devices. The device was placed inside a nitrogen cryostat and photoexcited using a 633nm diode laser. Fig. S5 shows the short-circuit photocurrent at various temperatures. We find that the photocurrent of P3TEA:SF-PDI$_2$ device drops more significantly than P3TEA:PCBM, with almost no photocurrent measured at ~100K. This is consistent with the temperature dependence found for the spectroscopy data as described in the main text. Data for annealed P3HT:PCBM, MEH-PPV:PCBM, PTB7:PCBM, and Si (reproduced from Gao et al. Phys. Rev. Lett. **114,** (2015), Gommans et al. Appl. Phys. Lett. **87,** 1–3 (2005), Ebenhoch et al. Org. Electron. **22,** 62–68 (2015), and Bambakidis & Smith NASA (1968)) are shown for comparison. We note that all device photocurrents were measured at sufficiently low intensity to avoid a current drop due to bimolecular recombination (with the exception of PTB7:PCBM which was measured under simulated solar illumination).

1.5. Additional time-resolved photoluminescence spectroscopy

Fig. S6 shows time-resolved PL of P3TEA:SF-PDI$_2$ at various detection wavelengths. As discussed in the main text, this delayed emission is observed only for the non-fullerene blend and not for the pure polymer or fullerene blended films (see Fig. 4) and is attributed to regeneration of emissive singlet excitons or CTEs upon recombination events of free carriers. We find little difference in the PL decay at various detection wavelengths, thus confirming that the exciton and CTE states are indeed very similar in energy (therefore allowing reversible interconversions between these states).

**Section S2. Extended technical descriptions of electroabsorption**

In this section we will explain the origin of the electroabsorption (EA) signals that we used to study charge separation in greater detail. The EA signal is created by a Stark effect of the absorption spectrum of the organic semiconductor. The Stark effect can be thought of as the shifting of a localized state energy in the presence of an electric field (E). The change in the excited state energy, ΔU is given by (10.1021/acs.jpclett.7b01741):

$$\Delta U(E) \propto -\Delta\mu E - \frac{1}{2}\Delta p E^2$$

where Δμ is the change in dipole moment and Δp is the change in polarizability.

There are 3 types of EA signals described in our study, as illustrated in Fig. S7b-d below. Although they have different origins, they are all caused by a uniform Stark shift of the absorption spectrum and therefore have a first-derivative lineshape. The first EA signal is the quasi-steady state EA signal measured in a diode structure. As discussed in the main text, this signal is used as a reference for comparison with the transient EA signals that we observed in pump-probe (PP) and pump-push-probe (PPP) experiments. The key difference between the quasi-steady state EA signal and the other two transient EA signals is that it is created by the macroscopic electric fields induced by the applied bias across the two electrodes, while the others are caused by local electric fields between photoexcitations during charge separation (with/without push excitation).

The second EA signal is observed in pump-probe transient absorption (in Δ*T/T*). As described in the main text and in previous reports (see Ref. [2,17] of main text) due to the derivative spectral shape of the EA response, the rise of EA signal causes a spectral blue-shift near the optical gap that provides a spectral signature for charge separation. It is important to emphasize that, unlike EA measurements performed on diodes, the resulted EA response is induced by multiple randomly oriented 'dipolar' fields localized in the vicinity of each photogenerated electron-hole pair. Thus, although the total EA intensity is still $\propto \int |E|^2 \, dV$, its origin is completely different. In the macroscopic case, the average electron-hole distance can be calculated assuming a simple model where two sheets of charges of opposite sign move away from one-another due to an applied bias. However, this macroscopic model is not valid for localised dipole-like fields where there is no preferential directionality to charge separation in a bulk heterojunction structure (the donor-acceptor interfaces are randomly oriented and the mean macroscopic field is zero). As described in previous work (Ref. [2] of main text), this local EA signal can be modelled by a classical electrostatic representation of the microscopic charge separation process where the work done to move an electron to a distance r from the hole is proportional to -1/r (with the energy at infinite distance normalized to zero). As the electron-hole separation increases, the total energy stored in the electric field will tend asymptotically towards that of two independent charges. Therefore, the change in electrostatic energy induced by moving these charges further apart becomes small once the electron-hole pair are 'free' from their mutual attraction. The electro-absorption induced by a charge pair will therefore follow the same trend. This implies that the local EA response is specifically sensitive to the separation from bound to free charges (signal saturates at ~5 nm separation). This signal is thus a signature for the charge separation process. As discussed in the main text, this signal is observed for both fullerene and non-fullerene blends, consistent with the high photocurrent IQE measured for both devices, but arising at very different rates (~100 fs vs. ~100 ps) and with contrasting temperature dependence.

The third EA signal is observed in pump-push-probe transient absorption (in $\Delta(\Delta T/T)$). As we discussed in main text, this signal arises upon push-excitation of bound charge-transfer excitons (CTEs) which stay localised at the donor-acceptor heterojunction. Absorption of the push energy allows the interfacial electron-hole pairs to spatially separate across the heterojunction, creating a local field that results in a microscopic EA response as described above. It is important to note that this push-induced EA signal is only observed in the presence of CTEs (as in the non-fullerene blends), and is not observed either when we push excitons (as in pristine film) or free carriers (as in the fullerene blends where most photoexcitations quickly separate into free charges). Therefore, the push-induced EA signal provides a direct probe of the localised CTE population at the donor-acceptor interface.

It is known that an EA signal caused by a Stark effect can have a first or second derivative lineshape (see Bublitz and Boxer, 10.1146/annurev.physchem.48.1.213). In cases where the Stark effect induces a change of polarizability of a transition, the electric field interaction will create a dipole moment only in the direction of the field, regardless of the molecule's orientation. As a result the entire absorption spectrum shifts uniformly to a lower energy, leading to an EA spectrum in the form of the first derivative of the absorption spectrum.

**Section S3. Extended technical descriptions of pump-probe (PP) data**

3.1. Non-normalised transient absorption data and spectral analysis

Fig. S8a-c shows the non-normalised transient absorption (TA) data of pure P3TEA, P3TEA:SF-PDI$_2$ and P3TEA:PCBM thin-films at room temperature, photoexcited at 670 nm at fluence of 1.5 µJ cm$^{-2}$ per pulse. The normalised data is shown in Fig. 2 in main text. These TA results were taken using a white-light continuum (~550-800 nm) created in a visible NOPA using the output of a Ti-Sapphire laser (see Methods). This wavelength window is ideal for studying spectral evolution close to the optical gap of P3TEA systems (~700-750 nm). Detailed description of the TA spectral evolution is found in the main text. We note that, since the selected excitation energy is well below the absorption edge of SF-PDI$_2$ and PCBM, we preferentially create photoexcitations in the P3TEA donor polymers. Under the same excitation condition, we observe no TA signal above the noise limit for a pure acceptor sample. We do observe a TA response when we directly excite the acceptor material at 500 nm. However, the intensity of the resulted TA signal is very weak (<10$^{-5}$). As shown in Fig. S8d, the TA response of the pure SF-PDI$_2$ film is only ~2x10$^{-3}$ when excited at a fluence of ~30 µJ cm$^{-2}$ per pulse. Under the same excitation fluence and accounting for difference in absorbance, we obtain a peak TA response of ~2x10$^{-2}$ for P3TEA and the blended films. This indicates that the absorption cross-section of excited states (both excitons and polarons) in the acceptor molecule is about an order of magnitude lower than the donor polymer, and therefore their presence in the blended films do not significantly affect our analysis. This is further confirmed by the largely overlapping TA response of P3TEA:SF-PDI$_2$ at early times under excitation at 500 nm and 670 nm (see Fig. S8e; i.e. the resulted spectra should look very different if excitons and polarons in the acceptor have comparable absorption cross-section to those in the polymer). We note that similarly small TA absorption cross-section was also found for excited states in PCBM

(see 10.1002/adma.201400846). For this reason, excited states in the acceptor domain can be neglected in our analysis.

We also measured TA data of the same samples over a wider spectral window (~580-1400 nm) by probing with a white light continuum generated by focusing the fundamental output of a PHAROS laser (1030nm) in a YAG crystal (see Methods). The results are summarized in Fig. S9. The results between ~580-780 nm are largely similar to those measured with the other TA setup as shown above. Here we attempt to differentiate the TA response of singlet excitons and polarons in P3TEA polymer. Since excitons generally require a donor-acceptor heterojunction to separate into charges, we consider that the TA response measured in the pure P3TEA film is mostly due to excitons. Although there is no spectral evolution between ~580-780 nm and ~1100-1400 nm, there is slight spectral change between ~780-950 nm for the pure P3TEA film. We note that this spectral region overlaps with the exciton emission spectrum. This spectral change can be due to either: 1) decrease in stimulated emission as excitons move towards less emissive sites, or 2) a small population of excitons that do manage to separate at disordered regions, forming polarons (see Reid, O. G et. al, Charge Photogeneration in Neat Conjugated Polymers. Chem. Mater. 26, 2014). Nevertheless, excitons are expected to be the dominant species in the pure film and therefore the resulting TA response can be attributed mostly to excitons (see blue line in Fig. S9d).

As described in the main text, singlet excitons are the dominant species at early times (<2 ps) in P3TEA:SF-PDI$_2$, while at later times there exists an equilibrium of excitons, CTEs and free carriers. Note that both CTEs and free carriers have polaronic TA signatures. For this reason, it is difficult to isolate the polaron response in the non-fullerene blend. On the other hand, excitons are able to quickly separate into free carriers in P3TEA:PCBM, and charges do not have sufficient energy to regenerate excitons upon recombination. Therefore, the measured TA spectrum at ~2-20 ps for the fullerene blend can be attributed to polarons, with substantial contribution from electroabsorption (EA) that causes spectral blue-shift near the optical edge (see red line in Fig. S9d).

3.2. Ultrafast and temperature-independent charge separation in P3TEA:PCBM

Fig. S10a shows the TA spectrum of pure P3TEA film at various temperatures (photoexcited at 670 nm at a fluence of ~1.5 µJ cm$^{-2}$). We observe a small spectral red-shift with reducing temperature, which is attributed to lowering of the optical gap. We observe little change in spectral shape with increasing time delay at all studied temperatures. Fig. S10b,c show the TA data of P3TEA:SF-PDI$_2$ and P3TEA:PCBM films studied at the same conditions. As described in the main text and Fig. 2d, the TA spectrum for the non-fullerene blend largely overlaps with the pure polymer film at early times (before ~2 ps), thus indicating that only excitons are present. Charges are able to form at later time, as evident by the spectral blue-shift caused by a growth in electroabsorption (EA) signal, but only when there is sufficient thermal energy (takes ~100 ps to complete at room temperature). For the fullerene blend, there is clear spectral blue-shift with respect to the pure P3TEA, thus indicates the presence of a significant EA signal even at a very early time scale (~0.2-1ps) and at cryogenic temperature (22K). This result is consistent with previous reports that show ultrafast and temperature-independent charge separation in fullerene blends (see Ref. [2,42] of main text).

3.3. Triplet recombination at nanosecond timescale in P3TEA:PCBM

Bimolecular electron-hole encounters at the heterojunction will generate thermalized CTEs, which can either have spin-singlet or spin-triplet characteristics according to spin-statistics (Ref. [9] of main text). As shown in Fig. S11, pump-probe transient absorption measurements show no evidence for triplet exciton formation in P3TEA:SF-PDI$_2$ on any timescale (top of panel a). In contrast, for the P3TEA:PCBM blend, our results show clear spectral red shift on nanosecond timescales that can be attributed to the formation of long-lived triplet excitons via bimolecular recombination (bottom of panel a). Panel b shows the decay of charges and the concomitant rise of triplets in the fullerene blend with the inset showing the triplet and charge spectra extracted. The triplet response shows overlap between the triplet photoinduced absorption signal and the 0-1 ground state transition at ~600-650nm. We speculate that the close energy alignment of the S$_1$ and CTE energy level suppresses triplet formation in the non-fullerene system, thereby allowing lower non-radiative recombination losses compared to the fullerene system.

**Section S4. Extended technical descriptions of pump-push-probe (PPP) data**

In this section we will explain the different components of the pump-push-probe (PPP) signal in greater detail. Figure S10a shows the PPP spectra of the P3TEA:SF-PDI$_2$ blend at pump-push delays from 0.8 to 800 ps. As mentioned in the main text, the signal primarily consists of two components: At long pump-push delays the spectrum is dominated by a derivative-like spectrum that can be attributed to an electroabsorption signal related to bound CTE states. The discussion in the main text is focused on this component, as it provides key insights into the charge separation dynamics investigated here. Here we describe the origin of the other component, which dominates at short pump-push delays.

Figure S12b compares this PPP signal at short push-probe delays with the inverted and scaled pump-probe (PP) signal measured at the same pump-probe delay. They match very well, which can be explained as follows: The push reduces the intensity of the transient absorption signal without significant changes in the spectrum. When calculating the difference in absorption, $\Delta(\Delta(T/T)) = (\Delta(T/T))_{push} - (\Delta(T/T))_{no\ push}$, this results in a signal that closely resembles the inverted PP spectrum. This reduction in $\Delta(T/T)$ signal indicates an overall reduction in excited state population. This is likely due to increased exciton-exciton annihilation of high-energy excitons formed by absorbing the push pulse (see Martini et al, Phys. Rev. B 2004, 69 (3), 1–12.), which also reduces the population of any states that form from the initial singlet excitons at later times. Fig. S12c shows the effect of temperature on the long-lived PPP signal in P3TEA:SF-PDI$_2$.

If this signal results from singlet excitons absorbing the push, we expect a similar signal in the pristine P3TEA film where no CTE states are formed. As shown in Fig. S13a, this is indeed the case. The spectra at 0.2 ps and 200 ps pump-push delay are similar and no derivative-like signal is present.

Since we know the origin and the shape of the spectrum that results from absorption by singlet excitons, we can remove this component and isolate the CTE response. We achieve this by subtracting the inverted pump-probe signal, scaled to minimise the signal around 900nm. We note that at very long pump-push delays (e.g, 800 ps) there is no contribution from the singlet annihilation signal anymore. Additional PPP results for P3TEA:PCBM are shown in Fig. S13b,c, and additional PPP results for P3TEA:FTTB-PDI$_4$, P3TAE:SF-PDI$_2$ and PffBT2T-TT:O-IDTBR are shown in Fig. S14.

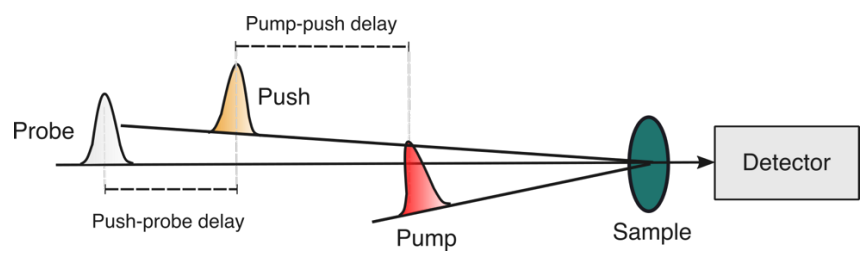

**Fig. S1**. Schematic illustration of time-resolved optical technique used in this work.

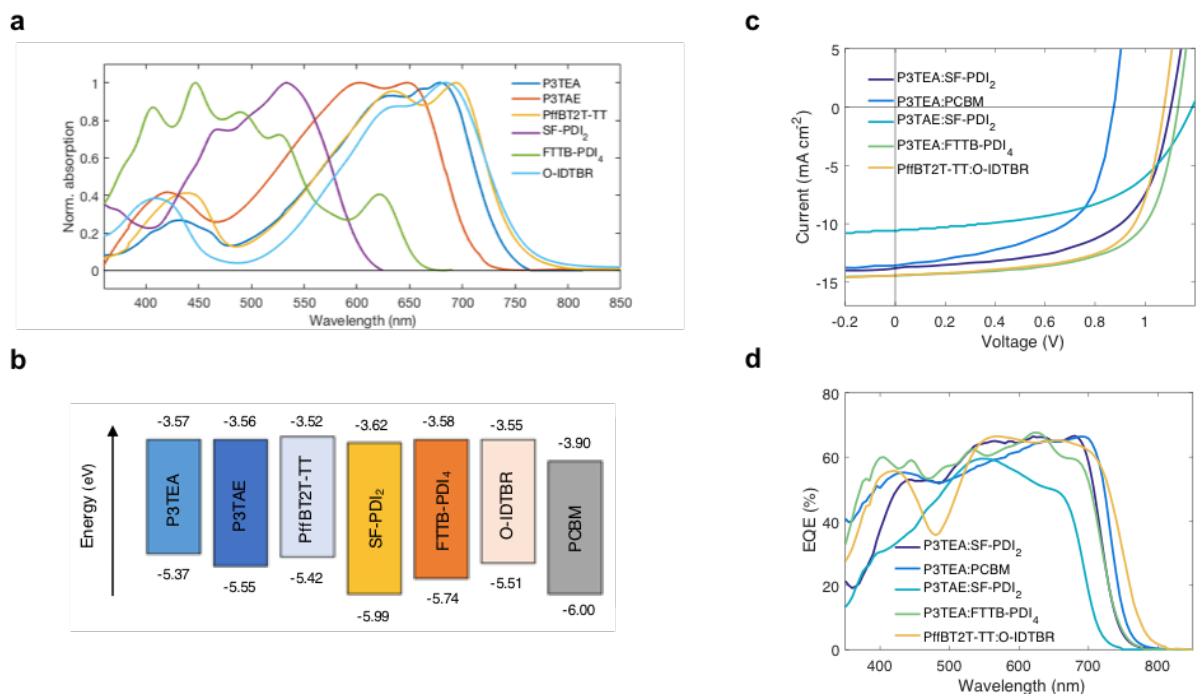

| | *$E_{gap}/q$ (V) | $V_{oc}$ (V) | $J_{sc}$ (mA cm$^{-2}$) | FF (%) | PCE (%) | *$E_{gap}/q - V_{oc}$ (V) | $EQE_{EL}$ | IQE (%) |
|---|---|---|---|---|---|---|---|---|
| **P3TEA:SF-PDI$_2$** | 1.72 | 1.11 | 13.27 | 64.3 | 9.5 | 0.61 | 5.0 x 10$^{-5}$ | ~85-90 (Ref. 27) |
| **P3TEA:PCBM** | 1.72 | 0.89 | 12.41 | 69.6 | 7.5 | 0.83 | 1.0 x 10$^{-6}$ | - |
| **P3TAE:SF-PDI$_2$** | 1.74 | 1.19 | 10.98 | 55.0 | 7.12 | 0.55 | 4.5 x 10$^{-4}$ | - |
| **P3TEA:FTTB-PDI$_4$** | 1.72 | 1.13 | 13.89 | 65.9 | 10.58 | 0.59 | 1.0 x 10$^{-4}$ | ~85-90 (Ref. 35) |
| **PffBT2T-TT:O-IDTBR** | 1.63 | 1.08 | 14.32 | 67.0 | 10.4 | 0.55 | 1.0 x 10$^{-4}$ | - |

**Fig. S2.** Material and device characterization. (a) Normalised UV-Vis data of materials involved in this study, all measured in thin-films. (b) Energy levels estimated from cyclic voltammetry. (c) Device J-V curves taken under simulated AM 1.5G sunlight. (d) External quantum efficiencies (EQE) at short-circuit condition. Device parameters are summarised in the table. *Optical gaps were determined from crossing of absorption/emission spectra.

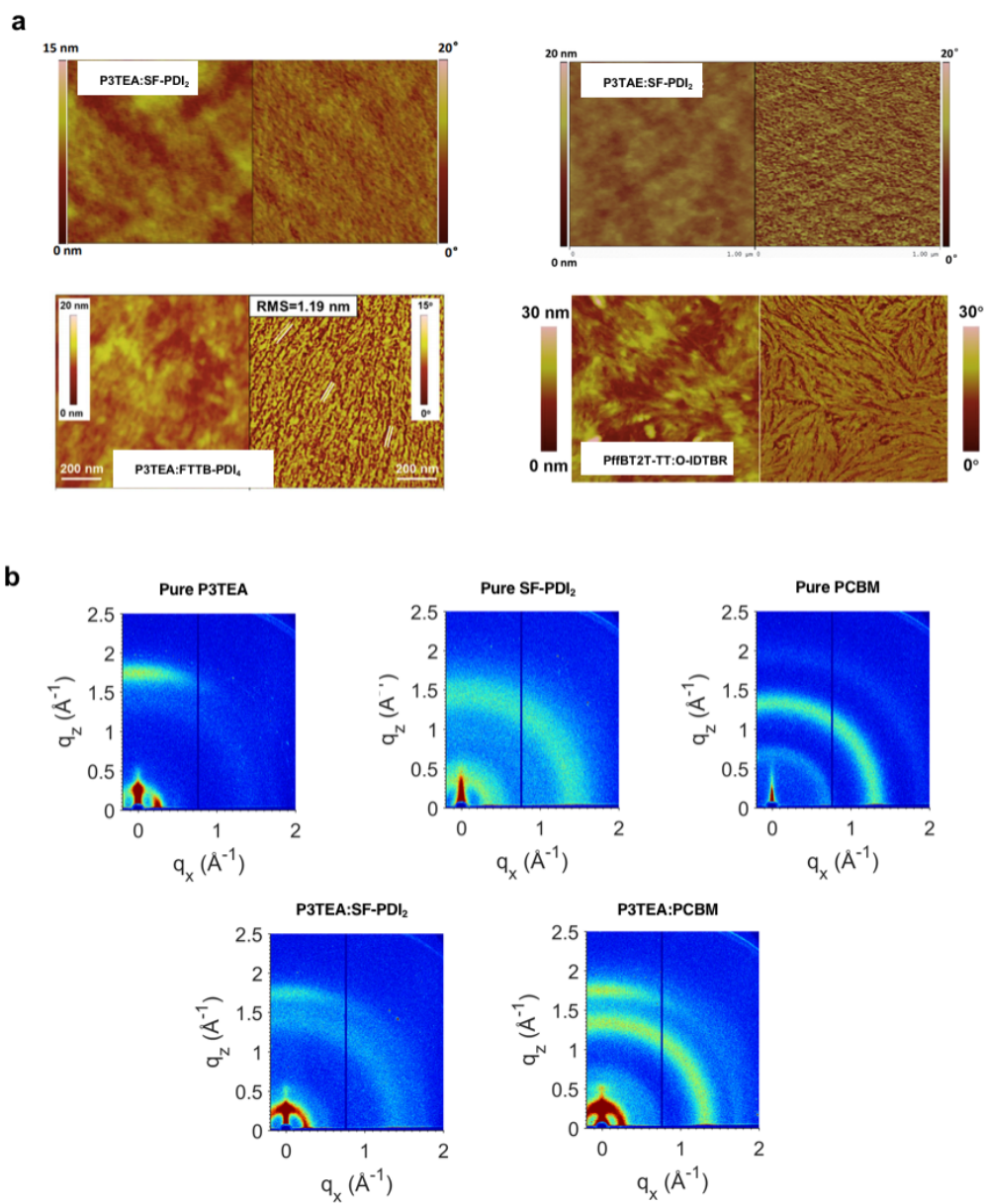

**Fig. S3.** (a) Atomic force microscopy (AFM) data showing uniform and smooth surface without clear phase separation in the non-fullerene blends. (b) Grazing-incidence wide-angle X-ray scattering of pure P3TEA, pure SF-PDI$_2$, pure PCBM, P3TEA:SF-PDI$_2$ and P3TEA:PCBM.

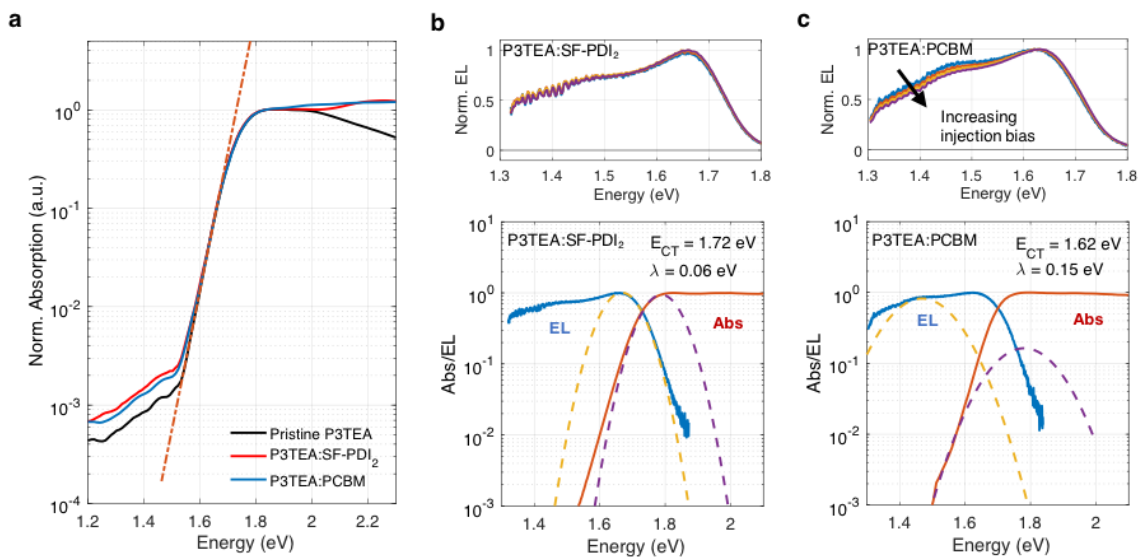

**Fig. S4.** Sub-bandgap absorption and emission. (a) Photo-thermal deflection spectroscopy (PDS) data of pristine P3TEA, P3TEA:SF-PDI$_2$ and P3TEA:PCBM films. The steep absorption edges show small Urbach energies of ~27 meV as indicated by the dashed line. (b,c) Electroluminescence (EL) spectrum of P3TEA:SF-PDI$_2$ (b) and P3TEA:PCBM (c) at various injection biases (between 1.5-3V) and simultaneous fitting of absorption edge and EL spectra with CTE energy ($E_{CT}$) and renormalization energy ($\lambda$) as fitting parameters, as described in Section S1.3.

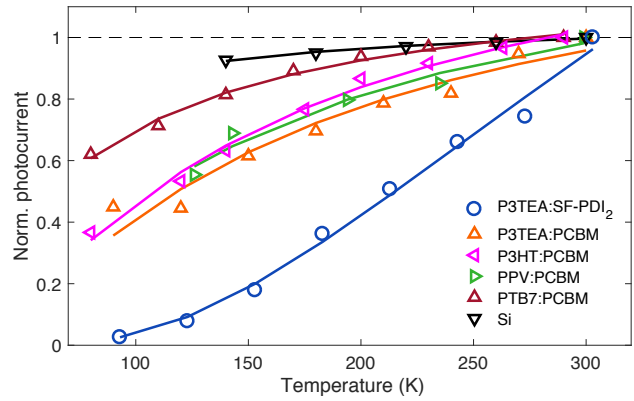

**Fig. S5.** Temperature dependence of device photocurrent. The device was placed inside a nitrogen cryostat and photoexcited using a 633nm diode laser. Data for annealed P3HT:PCBM, MEH-PPV:PCBM, PTB7:PCBM, and Si (reproduced from Gao et al. Phys. Rev. Lett. 114, (2015), Gommans et al. Appl. Phys. Lett. 87, 1–3 (2005), Ebenhoch et al. Org. Electron. 22, 62–68 (2015), and Bambakidis & Smith NASA (1968)) are shown for comparison.

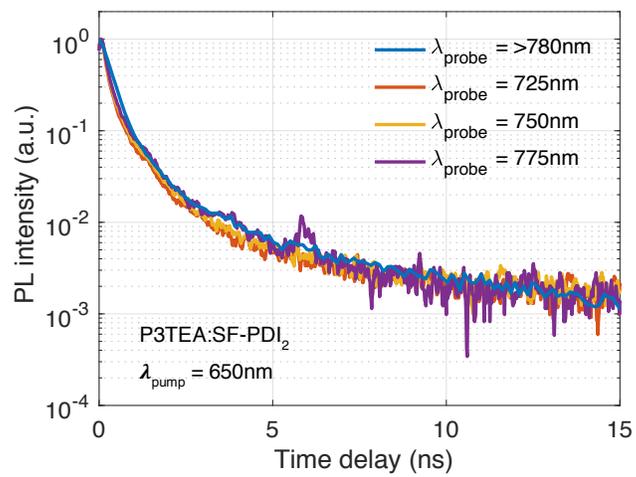

**Fig. S6.** Time-resolved PL decay of P3TEA:SF-PDI$_2$ thin-film photoexcited at 650 nm and probed at various wavelengths, showing negligible difference in decay lifetimes.

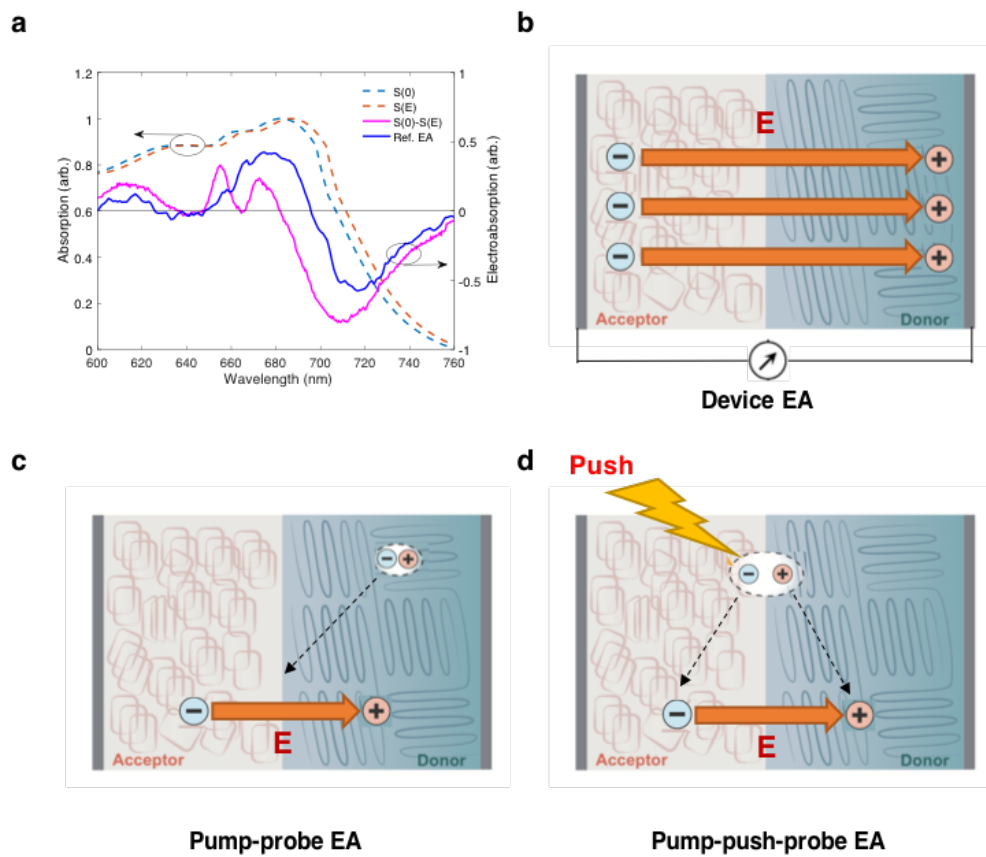

**Fig. S7**. (a) Stark shift of the absorption spectrum (S) due to an electric field E and the resulting electroabsorption response calculated from the difference between the shifted and unshifted spectra. The EA amplitude is proportional to $|E|^2$. The blue line shows the quasi-steady state EA measured in a diode structure for comparison. (b,c,d) Schematic illustrations of the 3 types of EA signals described in this study: EA signal caused by an external electric field (b), EA signal caused by photogenerated charges separating at the interface (c), and EA signal caused by charges initially bound at the interface that separate after interacting with the push pulse (d). Note that only a single donor-acceptor heterojunction is illustrated for simplicity. In reality the donor and acceptor are largely intermixed (bulk-heterojunction) with nanoscale domains, and the heterojunctions are randomly oriented.

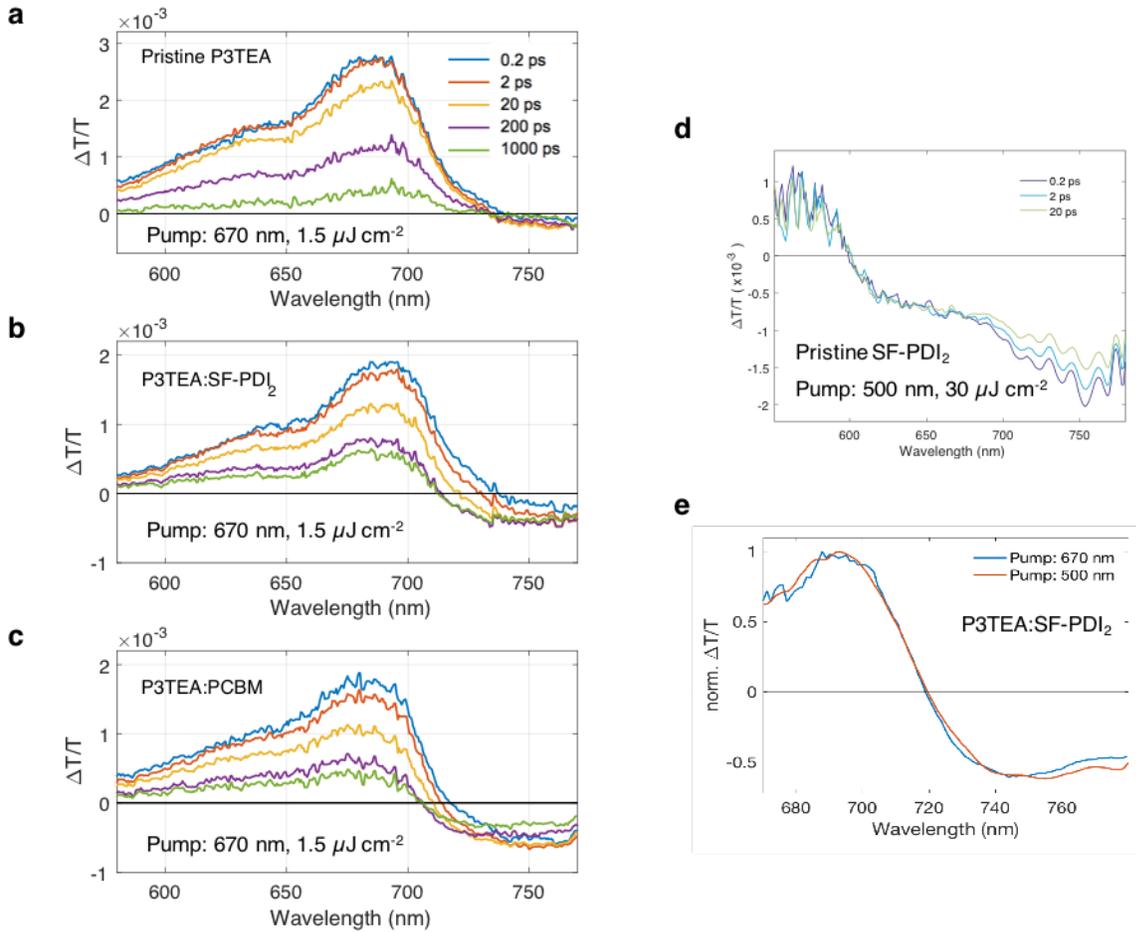

**Fig. S8.** (a,b,c) Un-normalised pump-probe transient absorption (TA) data of pristine P3TEA, P3TEA:SF-PDI$_2$ and P3TEA:PCBM. Normalised plots are shown in Fig. 2 in main text. (d) TA data of pristine SF-PDI$_2$ film, showing much weaker signal compared to samples containing P3TEA under same excitation density. We estimate the absorption cross-section of excited states in the acceptor is about an order of magnitude weaker compared to those in the donor polymer (similar values are also found for PCBM). (e) Normalised TA spectrum of P3TEA:SF-PDI$_2$ under excitation at 500 and 670 nm. The absence of a significant spectral evolution justifies neglecting the excited states in the acceptor domain when analysing the TA spectral evolution at this wavelength window.

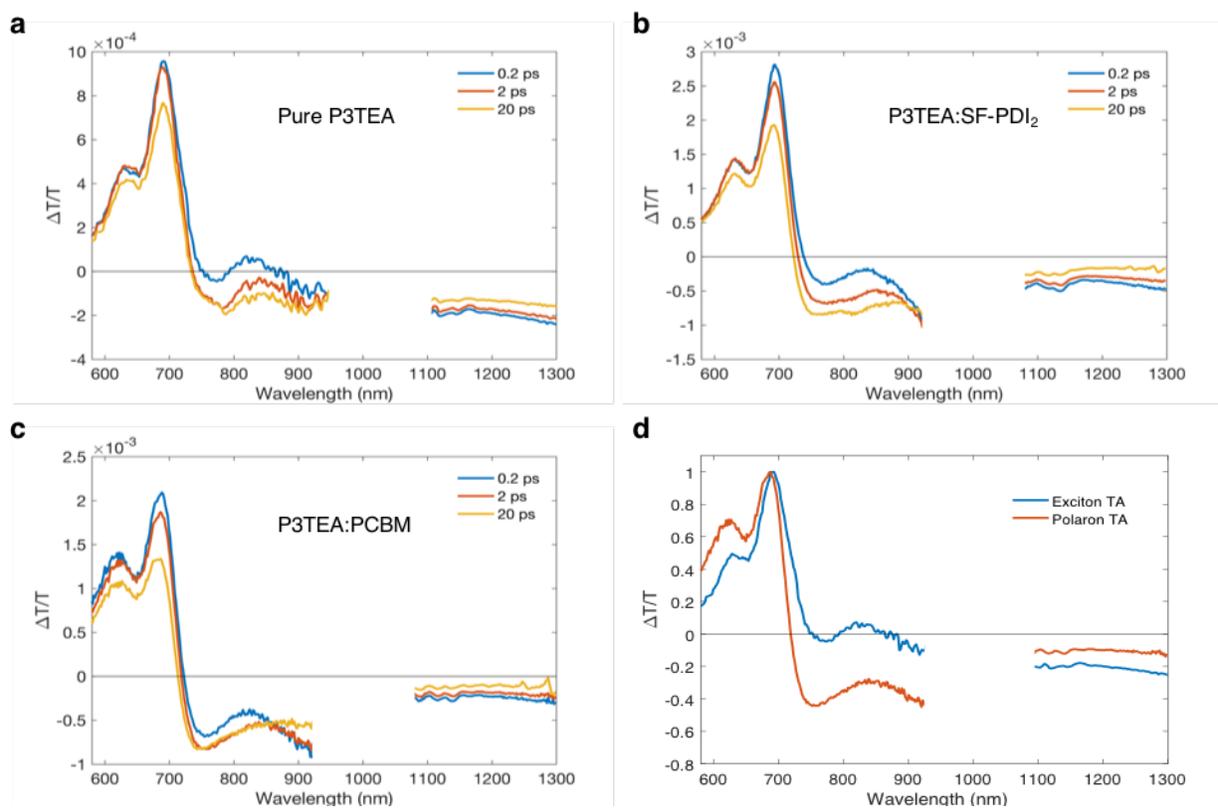

**Fig. S9.** (a,b,c) Un-normalised pump-probe transient absorption (TA) data of pristine P3TEA, P3TEA:SF-PDI$_2$ and P3TEA:PCBM over wide spectral window, excited at 670 nm at ~2 µJ cm$^{-2}$ per pulse. (d) Estimated TA spectrum of P3TEA singlet excitons and polarons (normalised to the peak signal at ~690 nm).

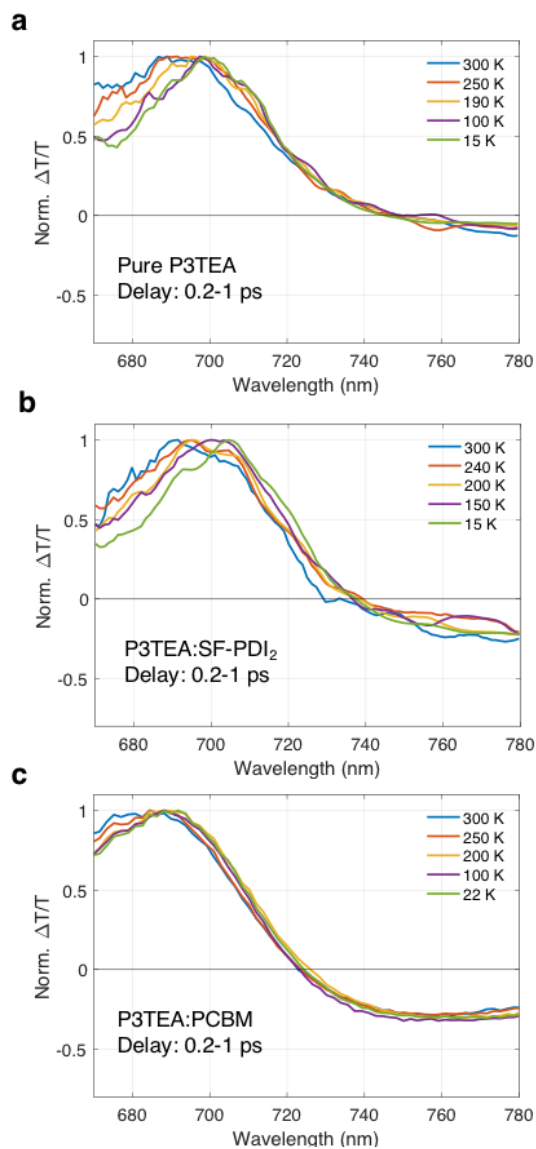

**Fig. S10.** Early-time (0.2-1ps) pump-probe (PP) transient absorption data of (a) P3TEA, (b) P3TEA:SF-PDI$_2$ and (c) P3TEA:PCBM, all excited at 670nm at a fluence of 1.5 µJ cm$^{-2}$ per pulse, measured at various temperatures. As shown in (a), the pristine P3TEA data show a slight red-shift of the peak ground state bleach (GSB) signal with reducing temperature, which can be explained by lowering of the optical gap. Similar feature is observed in the blend samples upon cooling. (b) TA data for P3TEA:SF-PDI$_2$ measured at this time range is very similar to the pristine polymer, indicating that excitons make up most of the excited states (charge separation is yet to happen). The small photoinduced absorption (PA) that emerges at ~760nm is likely due to polaron absorption following formation of bound CTEs. (c) For P3TEA:PCBM, the PA signal is even stronger, and the whole spectrum is already blue-shifted with respect to the pure P3TEA film from very early times 0.2-1 ps (crossing zero at ~720nm compared to ~740nm for the other 2 films). As described in main text, this blue-shift is due to emergence of the EA signal (negative from ~710nm onward), thus indicating ultrafast charge separation. Furthermore, this ultrafast separation process is unaffected by temperature (i.e. largely blue-shifted spectrum even at 22K).

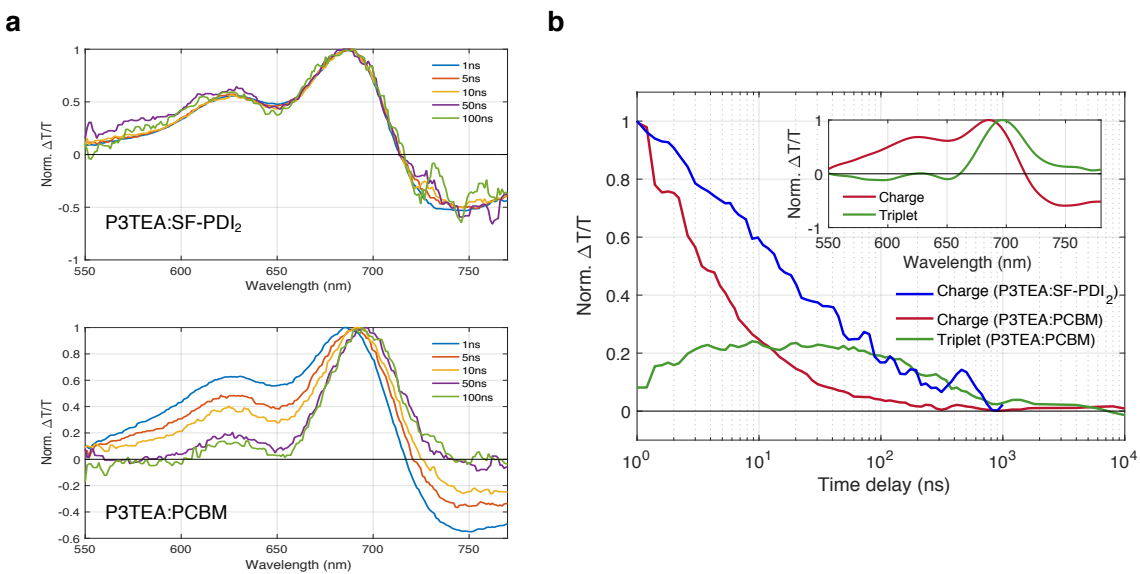

**Fig. S11.** Nanosecond pump-probe transient absorption (TA) spectroscopy. (a) TA data showing clear spectral shifting on nanosecond timescale due to formation of triplet excitons in P3TEA:PCBM but not in P3TEA:SF-PDI$_2$ (see Section S3.3). (b) Charge and triplet exciton signals in P3TEA:PCBM and P3TEA:SF-PDI$_2$ extracted via numerical spectral analysis of TA data, showing that in the P3TEA:PCBM blend charges recombine to form long-lived (~1μs) triplet excitons. No triplet formation was observed in the P3TEA:SF-PDI$_2$ blend. The inset shows the triplet and charge spectra extracted for P3TEA:PCBM, with the triplet response caused by the overlap between the triplet photoinduced absorption signal and the 0-1 ground state transition at ~600-650nm.

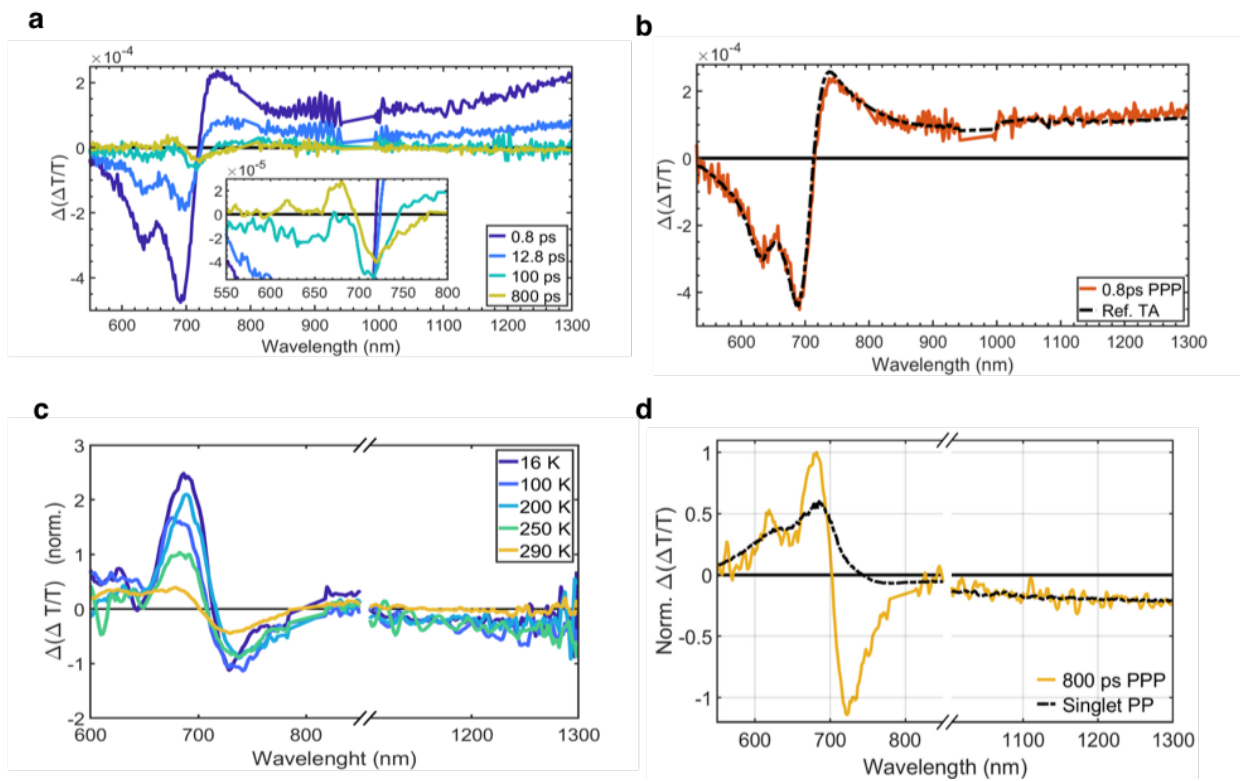

**Fig. S12.** Additional pump-push-probe (PPP) data of P3TEA:SF-PDI$_2$. (a) PPP signal at pump-push delays from 0.8 to 800 ps, integrated over push-probe delays of 2 to 5ps. The signal is normalised to compensate for changes in the overall excited states population absorbing the push pulse (divided by the pump-probe photoinduced absorption signal near the push wavelength). It therefore shows how the relative contribution of different sub-populations changes. Two domains can be distinguished; At long push delays a derivative-like feature dominates, magnified in the inset. At short push delays the signal corresponds to an inverted pump-probe signal. Subtraction of this 'exciton annihilation' signal reveals the derivative-like electroabsorption signal at long time delays (>50 ns; Figure 3b). (b) Comparison of this PPP signal at short pump-push delays (0.8 ps, with a push-probe delay integrated over 7 to 8 ps) with the inverted and scaled pump-probe data of the same sample at the same pump-probe delay. The $\Delta(\Delta T/T)$ signal is caused by a reduction in the $\Delta T/T$ signal, showing an overall decrease in excited state population. (c) Temperature dependence of pump-push-probe signal of the P3TEA:SF-PDI$_2$ film (averaged over pump-push delays of 50-100 ps and push-probe delays 0.4-0.5 ps), normalized by the PPP signal in the NIR region at 0.8ps. The signal is decreasing with increasing temperature, particularly between 250K and 290K, where it is reduced by more than half. (d) Comparison of PPP signal at 800ps with the singlet exciton PP signal taken from a pristine polymer sample. The agreement of the signals in the NIR indicates that at longer pump-push delay times the push can regenerate singlet excitons.

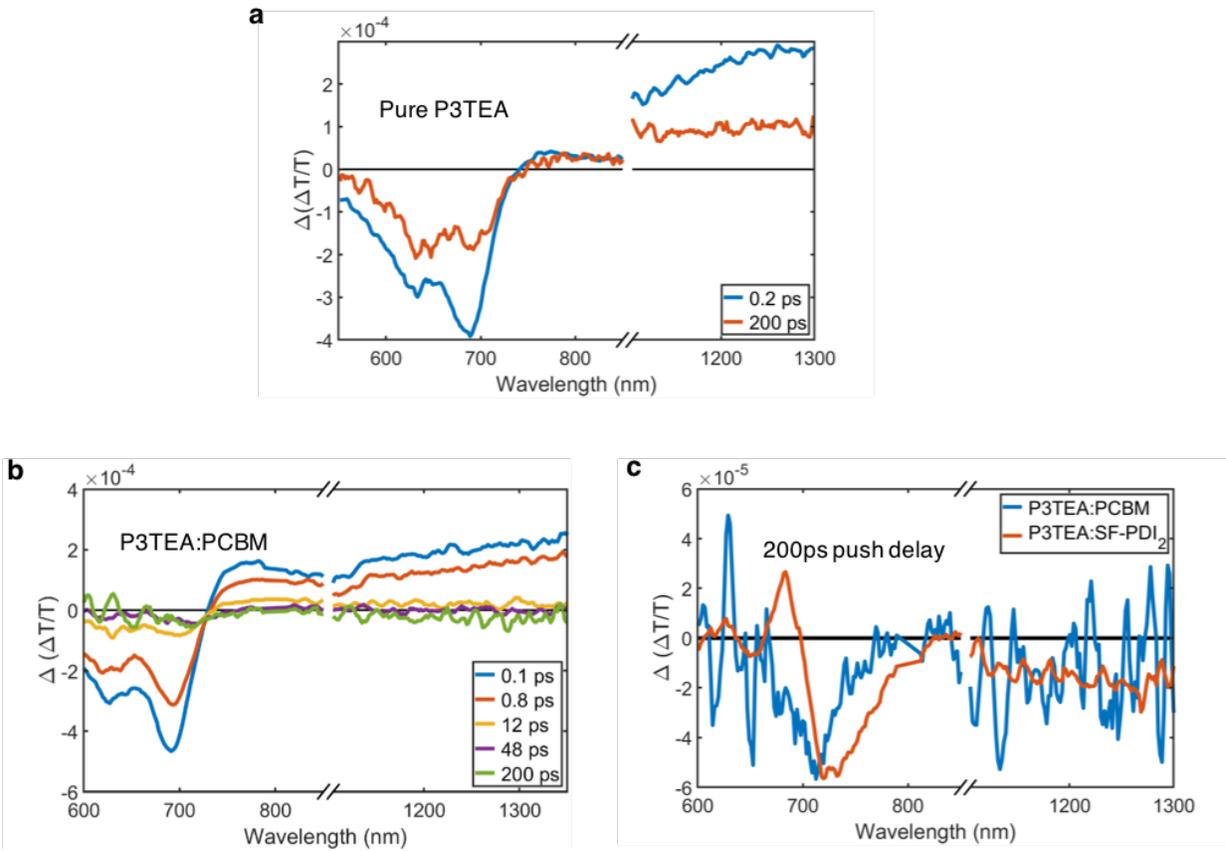

**Fig. S13.** (a) Pump-push-probe data of pristine P3TEA film (normalised and integrated as in Fig. S10). At all pump-push delays the signal resembles the inverted pump-probe signal, without a derivative-like signal emerging. (b) Pump-push-probe spectra of the P3TEA:PCBM film at various pump-push delays (normalised and integrated as in Fig. S10). At all pump-push delays the signal resembles the inverted pump-probe signal, without a derivative-like signal emerging. (b) Comparison of PPP signal of fullerene-based and non-fullerene films at long pump-push delays (200 ps, with a push-probe delay integrated over 1 to 2 ps). The non-fullerene blend shows the characteristic derivative-like line shape, while the fullerene-based blend does not.

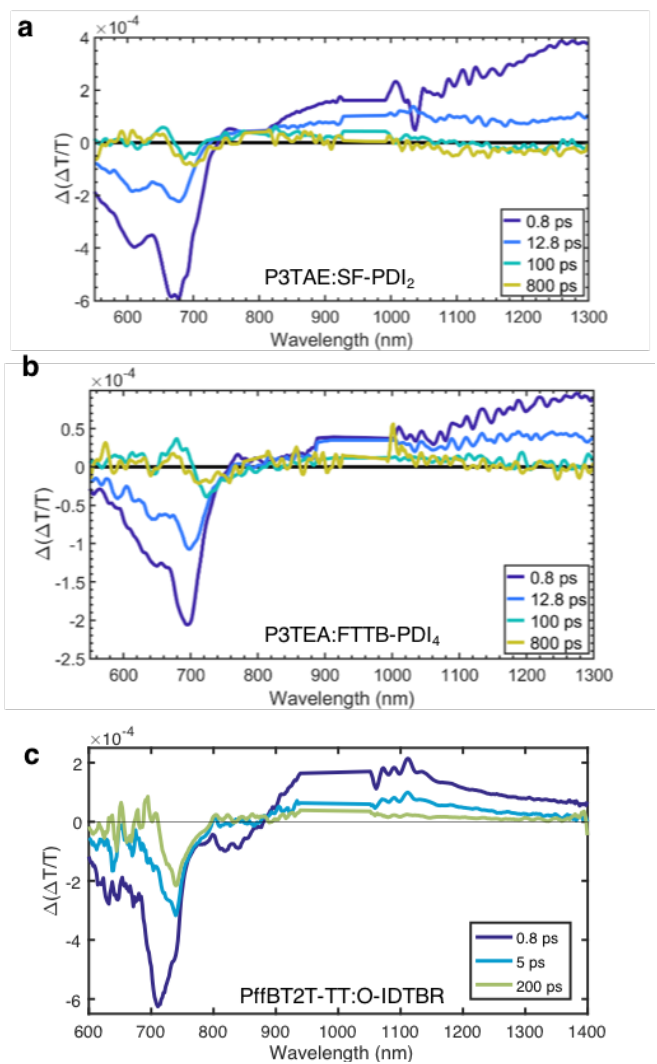

**Fig. S14.** Additional pump-push-probe data of P3TAE:SF-PDI$_2$, P3TEA:FTTB-PDI$_4$ and PffBT2T-TT:O-IDTBR films. (a,b) Data was integrated over push-probe delays of 1 to 2ps and normalised as in Fig. S10-11. (c) Data was integrated between 0.5 to 1ps push-probe delays. All blends show similar behaviour: an initial signal that corresponds to an overall reduction of excited states (inverse of ΔT/T signal observed in pristine polymer film), and a longer time signal with a derivative shape near the band edge which corresponds to separation of bound charge-transfer state (CTE) upon push excitation. As in the system discussed in the main text, this derivative-like feature is still present 800ps after the pump, thus indicating a quasi-equilibrium of CTEs and free charges.